\documentclass[showpacs,preprintnumbers,amsmath,amssymb,nofootinbib]{revtex4}
\usepackage{graphicx}
\usepackage{epsfig}
\usepackage{bm}
\usepackage{amsfonts}

\newcommand{\eq}{\begin{eqnarray}}
\newcommand{\eqx}{\end{eqnarray}}
\newcommand{\ba}{\begin{equation}}
\newcommand{\ea}{\end{equation}}
\newcommand{\f}[2]{\frac{#1}{#2}}

\newcommand{\n}{\nonumber \\}

\newcommand{\dl}{\delta}

\newcommand{\Dl}{\Delta}

\newcommand{\bit}{\begin{itemize}} 
\newcommand{\eit}{\end{itemize}}

\def\la{\label}

\def\bi{\bibitem}

\def\th{\theta}
\def\Del{\Delta}

\def\al{\alpha}

\def\be{\beta}

\def\va{\varphi}

\begin{document}

\title{From ``Dirac combs'' to Fourier-positivity}

\author{Bertrand G. Giraud }
\author{ Robi Peschanski }
\affiliation{Institut de Physique Th\'eorique,\\
CEA, IPhT, F-91191 Gif-sur-Yvette, France\\
CNRS, URA 2306 }
\email{bertrand.giraud@cea.fr; robi.peschanski@cea.fr}

\today

\begin{abstract}
Motivated by various problems in physics and applied mathematics, we look for
constraints and properties of real Fourier-positive functions, i.e. with
 positive Fourier
transforms.
Properties of the ``Dirac comb'' distribution and of its tensor products in
 higher dimensions
lead to
Poisson resummation, allowing for a useful approximation formula of a
 Fourier transform in
terms of a
limited number of terms. A connection with the Bochner theorem on positive
 definiteness of
Fourier-positive functions is discussed. As a practical application, we find
 simple and rapid
analytic algorithms for checking Fourier-positivity in 1- and (radial)
 2-dimensions among a
large
variety of real positive functions.
This may provide a step towards a classification of positive
 positive-definite
functions.
\end{abstract} 
\maketitle

\section{introduction}
\la{introd}

We call ``Fourier-positivity''  a property of a real positive function whose
Fourier transform is itself positive. Besides of a purely mathematical 
interest \cite{maths},
 the study of such pairs of functions is motivated  by applications in
 physics and applied mathematics \cite{applications}. In physics,
 typically, both functions correspond to observables, i.e. measurable
{\it a priori} positive
 quantities. Well-known examples
exist in one- and especially  two-dimensional cases. Let us quote for 
instance the
 Fourier-Bessel transform, i.e. radial version of the 2-dimensional 
 Fourier transform, relating the gluon and dipole distributions 
\cite{kovchegov}
 inside a hadron in the framework of Quantum Chromodynamics of  
strong particle interactions.

There exists a fundamental property characterizing Fourier positivity, which
uses the Bochner theorem \cite{bochner}: ``Fourier-positivity'' of a real 
function $\psi(\vec r)$ is equivalent to the statement that $\psi$ is not
only
positive but also positive-definite. ``Positive-definiteness''  means that
for any set of positions, $\{{\vec r_i}, i=1,...,n\}$, the $n \times n$
matrix
$\mathbb M$ with elements $\psi(\vec{r_i}\!-\!\vec{r_j})$ is positive
definite, i.e. 
\ba
\sum_{i,j=1}^{n}u_i\ \psi(\vec{r_i}\!-\!\vec{r_j})\ u_j\ge 0\quad 
\forall \vec u,\quad 
\forall n \in {\mathbb N}\ . 
\la{positive}
\ea
In other terms, the lowest eigenvalue of the matrix $\mathbb M$
 remains positive for all $\vec u$ and all values of $n.$

The Bochner theorem with its applications appears to be still the major
tool in 
the domain. However, to our knowledge, there does not yet exist a
mathematical
 classification of Fourier-positive  functions which, for instance,
could allow
for an appropriate parametrization for model building. Testing 
positive-definiteness  \eqref{positive} cannot be done concretely, due
to the 
generality of the constraints. Conversely, numerically computing 
Fourier transforms for checking Fourier-positivity is obviously 
possible, but it does not give general or analytical, easy means to select 
{\it a priori} appropriate Fourier-positive sets of functions.

Our approach is to find new constraints of Fourier-positivity allowing for 
simple and efficient selection rules of functions with positive Fourier
transform
 $\va$, given 
the positive input $\psi.$  We thus consider a pair of real even functions
on 
$d$-dimensional real
vector spaces, $\psi(\vec r)$ and 
$\varphi(\vec s),$ which are taken to be $d$-dimensional Fourier transforms
one from the other. 
\eq
\varphi(\vec s) &\equiv & \frac {1}{(2\pi)^{d/2}}\, \int_{{\mathbb R}^d}
d\vec r\ 
e^{i\vec s\cdot\vec r}\ \psi(\vec r)\ ,
\n \la{phi}
\n
\psi(\vec r) &\equiv & \frac {1}{(2\pi)^{d/2}}\, 
\int_{{\mathbb R}^d}
d\vec s\ e^{-i\vec s\cdot\vec r}\ 
\varphi(\vec s)\ .
\la{psi}
\eqx

We have already performed preliminary studies on this problem. 
In our initial paper \cite{gipe}, 
we examined the distribution of Fourier-positive functions among arbitrary 
combinations of a finite basis of Fourier eigenfunctions  using  the algebra
of
Hermite polynomials (in one dimension) and the algebra of Laguerre
polynomials 
(in radial two dimensions). The main outcome of this first study, using the 
Sturm algorithm on the number of polynomial zeros, is to reveal 
the rather intricate geometry of the manifold of solutions. In a second
study 
\cite{newfourier}, we derived generalized {\it sufficient} properties, based 
on an extension of convexity conditions by analytic continuation of $\psi$
into 
the complex plane and Jensen inequalities. However, the set of 
obtained constraints proved to be too 
weak to reliably check Fourier-positivity in the testing domain.
 
In the present work we propose a method providing a satisfactory detection 
of Fourier positivity, thoroughly tested for large sets of functions in one-
and radial two dimensions. It is based on a different approach from 
aforementioned studies, using 
remarkable properties  of ``Dirac comb'' mathematical distributions
 through Fourier transforms in any dimension.

 The plan of this paper 
is the following. In section II, we recall the definition and properties of
the
``Dirac comb'' distribution and of its tensor products,  leading to
 the well-known Poisson resummation formula. In section III, we then  derive 
a useful approximation formula of a Fourier transform in term of a finite 
(and limited) sum of data on the candidate Fourier-positive function. This
is 
shown, in section IV, to be equivalent to the application of the Bochner
theorem 
in a finite regular lattice of points of varying size. In section V we apply
 these results to large random sets of functions, either Fourier-positive or
not
 and show the  relevance of our method to select Fourier-positivity with an 
analytic evaluation of the Poisson summation. The final section VI is
devoted to 
a summary of results and prospects for a deeper understanding of 
Fourier-positivity.

\section{Dirac combs and the Poisson resummation formula}
\la{formul}
As we shall see, a key ingredient of our approach is to take advantage of
the Poisson resummation formulas, which can be easily derived from the 
so-called ``Dirac comb'' mathematical distribution,
\eq
\sum_{k \in 
{\mathbb Z}}\ \dl(r-k)\ &=&\ \sum_{k \in {\mathbb Z}}\ e^{2i\pi k r}\ ,
\n 
\n
\sum_{\vec k \in 
{\mathbb Z}^d}\ \dl^{(d)}(\vec r-\vec k)\ &=& \ 
\sum_{\vec k \in {\mathbb Z}^d}\ e^{2i\pi  \sum_{j=1}^d k_j r_j}\ ,
\la{comb}
\eqx
where the second line is its $d$-dimensional tensor product.

``Dirac combs''
are formally  invariant under Fourier transform, 
and also Fourier-positive. Indeed, as easily derived from  the definition 
\eqref{psi} inserted into the second line of \eqref{comb}
 one writes,
 \ba
 \int d\vec r\ 
e^{i\vec s\cdot\vec r}\ \sum_{\vec k \in 
{\mathbb Z}^d}\ \dl^{(d)}(\f{\vec r}{{2\pi}}-\vec k)\ = 
\ \ (2\pi)^d\ \sum_{\vec k \in {\mathbb Z}^d}\ 
e^{2i\pi\vec k\cdot{\vec s}}\ =\ \sum_{\vec k \in 
{\mathbb Z}^d}\ \dl^{(d)}(\f{\vec s}{{2\pi}}-\vec k)\ .
 \la{F-invariant}
 \ea
The connection between Dirac combs and Fourier-positive functions is obtained 
by considering the  ``characteristic function'' defined by:  
\ba
F(\vec\th,\vec r) \ \equiv\ \sum_{\vec k \in {\mathbb Z}^d}\ 
\psi( {{k_1r_1},\dots,{k_dr_d}})\ \ e^{i\sum_{j=1}^d k_j\th_j}
\ .
\la{iffd}
\ea
As we shall discuss in further sections the condition for $\psi$ to be 
Fourier-positive transfers the condition
to a positivity of $F(\vec\th,\vec r),$ namely,
\ba
F(\vec\th,\vec r)\ >\  0 \quad \forall\ \vec r,  \vec \th \ 
\in {\mathbb R^d\otimes [0,2\pi[^d} \ .
\la{conditiond}
\ea
In fact, noting from the second line of \eqref{psi} that, 
\ba
\psi(k_1r_1,\dots,k_dr_d)  = \ \frac {1}{(2\pi)^{d/2}}\, 
\int_{{\mathbb R}^d}
d\vec s\ e^{-i\sum_{j=1}^d k_jr_js_j}\ 
\varphi(\vec s)\  ,
\la{kphid}
\ea
we are able to use the $d$-dimensional ``Dirac comb'' relation \eqref{comb}
in order to rewrite the positivity condition  \eqref{conditiond} as,
\eq
F(\vec\th,\vec r) \ &=&\ \frac {1}{(2\pi)^{d/2}}\, 
\int_{{\mathbb R}^d}
d\vec s\ 
\varphi(\vec s)\ 
 \sum_{\vec k \in {\mathbb Z}^d} \ e^{-i\sum_{j=1}^d k_j(r_js_j-\th_j)}
\n
 \ &=&\ \frac {1}{(2\pi)^{d/2}}\, 
\int_{{\mathbb R}^d} 
d\vec s\ 
\varphi(\vec s)\ 
\left\{ \sum_{\vec k \in 
{\mathbb Z}^d}\Pi_{j=1}^d\dl\left(\f{r_j s_j -\th_j}{2\pi}-
 k_j\right)\right\}
\n
 &=&\ \f {(2\pi)^{d/2}}{\Pi_{j=1}^d| r_j|}\ \sum_{\vec k \in 
{\mathbb Z}^d}\varphi\left( 
{\f
{2\pi k_1 + \th_1}{r_1},\dots,\f
{2\pi k_d + \th_d}{r_d}}
\right)\ > \ 0\ \quad \forall\ \vec r,  \vec \th \ \in \mathbb R^d\otimes 
[0,2\pi[^d\  .
\la{iff2d}
\eqx
The equality of the two expressions \eqref{iffd} and \eqref{iff2d} 
of $F(\vec\th,\vec r)$ is nothing but a version of the $d$-dimensional 
Poisson resummation formula \cite{poisson}.

An interesting insight on the properties of  \eqref{iff2d}
is obtained by a change of variables $(\th_j,r_j) \Leftrightarrow (s_j,r_j)$ 
with 
\ba
s_j \equiv \f{\th_j}{r_j},\quad j=1,\dots,d\ ;\quad F(\vec\th,\vec r)
\Leftrightarrow 
F(\vec s,\vec r)\ .
\la{changevariables}
\ea
The positivity condition \eqref{conditiond} may thus be rewritten 
and renormalized in such a way as to read,
\ba
F(\vec s,\vec r)\ \equiv \sum_{\vec h \in {\mathbb Z}^d}
\varphi\left( s_1\!+\!\f{2\pi h_1}{r_1},
\dots,
s_d\!+\!\f{2\pi h_d}{r_d}\right)\ =\ \f {|r_1\dots r_d|}{(2\pi)^{d/2}} 
\sum_{\vec k \in {\mathbb Z}^d} 
\psi( {{k_1r_1},\dots,{k_dr_d}})\ e^{i\sum_{j=1}^d k_jr_js_j} \ >  0\, . 
\la{sum}
\ea
Under this form, the Poisson resummation formula \eqref{sum} allows 
for an interesting Fourier transform relation  which  can be 
qualitatively (and made quantitative in the next
section) outlined as follows,

\begin{itemize}
\item {\it Approximation of the right-hand side of Eq.(\ref{sum}).}  Assume
 that the
function 
$\psi(r_1,\dots,r_d)$ has a finite range $R$ for each of its arguments,
 namely that it
is negligible\footnote{It is assumed to be small enough even 
in a summation like in \eqref{sum}.} if any $|r_j|> R$.
Then one can choose a positive integer $K$ so that the right-hand summation
 can be
truncated 
into a finite number of terms. Indeed, define a parameter, $r_{\rm min} =
 R/K$, and 
consider only situations  where $|r_j| > r_{\rm min}$, $\forall j=1,\dots,d$.
 Clearly,
every $|k_j|$ becomes bounded by $K$, hence,
\ba
F(\vec s,\vec r)\  \simeq\ \f {|r_1\dots r_d|}{(2\pi)^{d/2}} 
\sum_{\vec k \in {\mathbb Z}^d\, ;\, |k_j| < K,\, \forall j} 
\psi( {{k_1 r_1},\dots,{k_d r_d}})\ \ e^{i\sum_{j=1}^d k_jr_js_j}.
\la{approxpsi}
\ea
Given $K$, there appears a $minimal$ value of each $r_j$ for practical
 calculations
if $K$ remains fixed.

\item {\it Approximation of the left-hand side of Eq.(\ref{sum}).} Assume
 also that
the function 
$\va(\vec s)$ has a finite range for each of its arguments, $|s_j|<S,\,
 \forall j$,
beyond which it is negligible.
Then one finds  that the left-hand summation can be limited to its first
 term 
$h_j=0\ ,j=1,\dots,d$. This only remaining significant term is just the
 Fourier
transform of $\psi,$ hence,
\ba
\sum_{\vec h\in {\mathbb Z}^d}\varphi\left( s_1\!+\!\f{2\pi h_1}{r_1},\dots ,
s_d\!+\!\f{2\pi h_d}{r_d}\right) \ \simeq \varphi\left( s_1,\dots ,s_d\right)
 \ , 
\la{approxphi}
\ea
This holds for some $maximal$ value of each $|r_j|$, depending on $S$. This
 maximum, 
$r_{\rm max}$, will be derived in the next section.
 
\end{itemize}

Let us comment these two approximations. Eq.\eqref{approxpsi} is meant to
obtain a good approximation of the characteristic function itself. 
Obviously, this expression
exhibits an approximation formula for a Fourier transform, and,
usually, a convergent result at the limit, $K \rightarrow \infty$, or, as
 well, 
$r_{\rm min} \rightarrow 0.$. This requires 
a priori $K$ to be large enough. However, our aim is to look for a good 
enough
approximation for a limited number of terms in \eqref{approxpsi}.  
In this case the  variable $r_{\rm min}$ plays the role of 
a ``resolution'' 
parameter, $\Del r = R/K$, on
the function $\psi$ allowing to
test its Fourier-positivity. The approximation described by Eq.
\eqref{approxphi} is
of different nature and is directly related to the properties of the Poisson 
resummation. Indeed, the trade of variables $k_i r_i$ into $s_i\!+\!\f{2\pi 
h_i}{r_i},$ typical of the Poisson resummation formula \eqref{sum}, allows for a
rapid decrease rate in the left-hand series, for small enough values of $r_i.$

By combining both approximations (\ref{approxpsi},\ref{approxphi}), one has the 
interesting approximation property, valid in a 
restricted domain for $\vec r$, of a Fourier transform by a finite sum, 
\ba
\varphi( s_1,\dots,
s_d )\ \simeq\ \f {|r_1\dots r_d|}{(2\pi)^{d/2}}\ \sum_{ k_j=1}^K \ 
\psi( {{k_1r_1},\dots,{k_dr_d}})\ e^{i\sum_{j=1}^d k_jr_js_j} 
\la{approxfourier}
\ea
allowing to check the positivity and other properties of $\va$ from a finite number of
values of $\psi$ for the variables $r_j$ in a given range,
$r_{\rm min}<|r_j|<r_{\rm max}$,
delimited by both a lower bound (related to the right-hand side approximation) and an
upper bound (related to the left-hand side approximation) induced from the Poisson
resummation
\eqref{sum}.

\section{Fourier transform $via$ Poisson resummation}
\la{proleg}

\subsection{The one-dimensional Fourier case}
\la{oned}
\medskip

Here we consider a conjugate pair of real even functions  $\psi(r),\va(s)$ of real
variables which are
Fourier-conjugated one with the other, 
see formulas
\eqref{psi} when $d\!=\!1.$ We look for 
the application of the characteristic function  $F(s,r)$ 
and the corresponding Poisson resummation formula \eqref{sum}
to the one-dimensional problem of Fourier positivity, using the approximation
scheme \eqref{approxfourier} outlined in the previous section.

Let us first write the Poisson resummation formula \eqref{sum} in 
the one-dimensional case.
\ba
F(s,r)\   =\ \sum_{h \in {\mathbb Z}}\varphi\left( s+\f{2\pi h}{r}\right)\ =
\ \f {|r|}
{\sqrt{2\pi}}\ \sum_{k \in {\mathbb Z}}\ \psi(kr)\ e^{ikrs}\ . 
\la{sum1}
\ea
Under this form, the  approximation properties leading to 
\eqref{approxfourier}
can be made quantitative as follows;

\begin{itemize}
\item {\it Approximation of the right-hand side summation in 
Eq.\eqref{sum1}.} 
Let us consider a fixed value, moderately large, of the positive integer
parameter $K$.
Consider a function $\psi(r)$ with a finite and non-zero range $R$, define the
parameter,
$r_{\rm min}=R/K$, and restrict $r$ to be larger than $r_{\rm min}$. This
gives,
\ba
F(s,r)\  \simeq\ \frac{|r|}{\sqrt{2 \pi}} \sum_{|k|\le K} \psi(kr)\ e^{ikrs}.
\la{approxpsi1}
\ea
This summation in \eqref{sum1} is thus
limited\footnote{We assume that the cut-off $R$ is strong enough to ensure
a fast convergence of the series \eqref{approxpsi1}.}
to just terms with $|k| \le K $ and looks like a discretized approximation,
with
step $r$, of the Fourier integral.

\item {\it Approximation of the left-hand side summation in 
Eq.\eqref{sum1}.}  Let us
assume a finite range $S$ of
the function $\va(s)$, Fourier partner of $\psi$, namely $\varphi(s)$ is
negligible
if $|s| > S$.

Besides an obvious condition, $|s|<S$, the condition for retaining the first
term,
$h=0$, as the only significant term in the left-hand side summation
\eqref{sum1},
is,
\ba
\left\vert s \pm \f{2\pi}{r}\right\vert\, > S,
\la{approxphi1}
\ea
since all other terms, with $|h|>1$, can then be neglected (provided one has
good
convergence properties). A bound on  $|r|$ is obtained as follows. Clearly,
whatever
the signs of $r$ and $s$, the condition \eqref{approxphi1} reduces to,
\ba
\left| |s|-\frac{2\pi}{|r|} \right| > S,
\ea
with two branches, depending on the relative values of $|s|$ and $2 \pi /|r|$,
\ba
|s|-\frac{2\pi}{|r|} > S, \ \ \ \ \ -|s|+\frac{2\pi}{|r|} > S.
\ea
Since $|s|<S$, the first branch is useless. The second one gives
$2\pi/|r|>2S$, because
$|s|$ may reach $S$. Accordingly, one finds the bound,
$|r|<r_{\rm max}\equiv \pi/S$.

By combining both approximations, one obtains the approximation property of a
Fourier transform by a finite sum, namely,
\ba
\varphi( s )\ \simeq\ \f {|r|}{\sqrt{2\pi}}\ \sum_{k=1}^K \ 
\psi(kr)\ e^{i krs}\ . 
\la{approxfourier1}
\ea
For \eqref{approxfourier1} to be valid, one has to choose an appropriate
range for
$|r|,$ namely,
\ba
\f{R}{K} \ <\ |r| \ <   \f{\pi}{S}\ . 
\la{range1}
\ea
This, in turn, requires the following condition on the truncation parameter,
$K$,
\ba
{K} \ >\   \f{RS}{\pi}\ ,
\la{limited1}
\ea 
the value of which is not too large for a pair of Fourier partners whose 
cut-offs are such that the product, $R\, S$, is a finite number.
\end{itemize}

\subsection{The radial two-dimensional Fourier case}
\la{proleg2}
The radial two-dimensional  problem considers $\psi(r)$ and
 $\va(s),$ a  pair of radial conjugate functions on ${\mathbb R}^+$, namely 
\eq
\varphi(s)\ &=&\ \frac {1}{2\pi}\, \int_{{\mathbb R}^2} d\vec r\ 
e^{i\vec s\cdot\vec r}\ \psi( r)\ =\ \int_{0}^{+\infty}\!\!\!\! rdr\ J_0(sr)
\  \psi(r)\, ,
\la{phiJ}\\
\psi(r)\ &=&\ \frac {1}{2\pi}\, 
\int_{{\mathbb R}^2}
d\vec s\ e^{-i\vec s\cdot\vec r}\ 
\varphi(s)\ =\ \int_{0}^{+\infty}\!\!\!\! sds\ J_0(rs)\  \varphi(s) 
\la{psiJ}\ , 
\eqx
where 
$r=\vert \vec r\vert$ and $s=\vert \vec s\vert$ denote here 
the radial 
variables of vectors in both conjugated 2-dimensional spaces.

Following the general formalism of section \ref{formul}, the two-dimensional 
Poisson resummation ensures the positivity of the 2-dimensional characteristic
function for a positive Fourier transform $\va(s),$ namely,
\ba
F(s_1,s_2,r) \ \equiv \ \  \f {r^2}{2\pi}\  \sum_{m,n\in 
{\mathbb Z}}\psi\left(r\sqrt{m^2+n^2}\right)\ e^{i(ms_1+ns_2)r}\ \ge \ 0
\quad \forall  r \in [0,\infty[, \quad \forall s_{1},s_2\ .
\la{F2d}
\ea
 
As we discuss now, the condition \eqref{F2d} happens
to furnish also a  condition for 
Fourier-positivity of the function $\psi(r)$ in the range where 
$F(s_1,s_2; r)$ gives a direct link to its Fourier transform
similar to the one-dimensional case. This comes again from a
2-dimensional Poisson summation formula relating $\psi(r)$ to $\va(s)$,
namely,
\ba
F(s_1,s_2, r) \ =\ \sum_{ h_{1},h_2 \in 
{\mathbb Z}}\varphi \left(
{\sqrt{(\f {2\pi} r h_1 + s_1 )^2+ 
(\f{2\pi} r h_2 + s_2)^2}}
\right) 
\ \ge \ 0\ .
\la{iff2d2}
\ea
\begin{itemize}
\item {\it Approximation of the summation in Eq.\eqref{F2d}.} 
Let us consider a fixed value of the variable $r.$ Then, we consider functions
$\psi$ with a finite range $R$.
It is easy to see that the summation \eqref{F2d} can be 
limited\footnote{We assume that the cut-off $R$ is strong enough to ensure
a fast convergence of the series\eqref{F2d}.}
to just terms with $|m|,|n| \le K \equiv [R/{r}]+1$ where $[R/{r}]$ is the 
integer part of $R/r\ .$
\ba
F(s_1,s_2, r)\ \simeq \ \frac{r^2}{2\pi} \sum_{m,n\in 
{\mathbb Z}\, ;\, |m|,|n| < K }\ \psi\left(r\sqrt{m^2+n^2}\right)\ 
e^{i(ms_1+ns_2)r}
\quad {\rm with} \ K{r}\ge R \ . 
\la{approxpsi2}
\ea

\item {\it Approximation of the summation in Eq.\eqref{iff2d2}.}  Consider
the case of a finite range $S$ of the  function $\va$, Fourier transform of
$\psi$. Namely, assume that  $\va(s)$ is negligible for $s=\sqrt{s_1^2+s_2^2}>S.$ Then, consider the set of
points, $\{s_1,s_2\}$, inside the circle with radius $S$, namely, $s_1^2+s^2<S$.
The conditions for retaining only the term with, $h_1=h_2=0$, in
\eqref{iff2d2} read,
\ba
\left(\pm \f {2\pi} r  + s_1 \right)^2 + (s_2)^2 > S^2 \ 
({\rm for}\ h_1=\pm 1,h_2=0)
\ \ \ {\rm and}\  \ \ 
(s_1)^2 + \left(\pm \f {2\pi} r + s_2\right)^2 > S^2 \ 
({\rm for}\ h_2=\pm 1,h_1=0) \ .
\la{approxphi2}
\ea
This is obtained if, $2\pi/r > 2 S$.
Indeed, under such a condition for $2\pi/r$, the points,
$\{\pm 2 \pi/r + s_1,s_2\}$, are
pushed out of the circle which confines $\{s_1,s_2\}$. And the same holds for
the
points $\{s_1,\pm 2\pi/r+s_2\}$.
Accordingly, the summation \eqref{iff2d2} can be reduced\footnote{We assume
again
that the cut-off $S$ is strong enough to ensure
a fast convergence.} to its simplest term, $\vec h=0$, i.e., the only
remaining
significant term. It is just the Fourier transform $\va(s)$ of $\psi.$ 

By combining both approximations, one obtains the approximation property of a
Fourier
transform by a finite sum, namely,
\ba
\varphi(s)\ \simeq\ \frac{r^2}{2\pi}\ \sum_{m,n \in {\mathbb Z}\, ;\,  
|m|,|n| < K} 
\psi\left(r\sqrt{m^2+n^2}\right)\ e^{i(ms_1+ns_2)r}\ . 
\la{approxfourier2}
\ea
For \eqref{approxfourier2} to be valid, one has to 
choose appropriate bounds for
$r,$
which are similar to those of the $1d$ case, namely,
\ba
\f{R}{K} \ <\ r \ <   \f{\pi}{S}\quad \Rightarrow\quad {K} \ >\   
\f{RS}{\pi}\ . 
\la{range2}
\ea
\end{itemize}

\section{From Poisson formula to Bochner positive-definiteness}
\la{poissonbochner}

Our aim in this section
is to exhibit  the connection of the positivity condition 
\eqref{conditiond} of the 
characteristic function  $F(\vec\th,\vec r)$
with the positive-definiteness of certain 
moment matrices \cite{akhiezer} related to the Bochner theorem
 \cite{bochner}. This in turn implies   the positivity of the lowest
eigenvalue
of these matrices (and thus also of the corresponding matrix 
determinants). For practical reasons, we shall focus the discussion on the
one-
 and radial 
two-dimensional cases, but the method is general.

\subsection*{The one-dimensional case}
\medskip

Let us recall the one-dimensional Poisson formula \eqref{sum1} and its 
positivity condition under the form,
\ba
F(\th,r) \ \equiv\ \sum_{k \in {\mathbb Z}} \psi(kr)\ \exp{(ik\th)}\ = \ 
\frac{\sqrt{2\pi}}{|r|}\ \sum_{k \in 
{\mathbb Z}}\varphi\left( \f{2\pi k+\th}r\right )\ \ge \ 0\ , \quad 
\forall\ (r,\th)
\in\ (\, ] -\infty, \infty[\, \otimes\, [0,2\pi[\, ) \ .
\la{iff}
\ea
As discussed in the preceding section, the positivity condition \eqref{iff}
happens to be an equivalent 
formulation of the Fourier-positivity of the input function $\psi(r)$. 
In the following we show that this positivity condition can be rephrased
in terms of a specific application of the  Bochner theorem \cite{bochner}.
 
In the present case, starting from Eq.\eqref{iff}, the 
problem\footnote{In the mathematics literature,
one may
refer to the  {\it problem of moments} \cite{akhiezer,shohat}, and
 more specifically in our case to the {\it trigonometric moment problem} 
\cite{akhiezer,cara,toeplitz}. The 
trigonometric moment problem  
can be expressed as follows: Find a bounded, positive function 
$F(\th)\ge 0,\ \th\in [0,2\pi]$ such that 
its trigonometric moments,
\ba
\mu_n \equiv  \f 1{2\pi}\int_0^{2\pi}\!\ e^{in\th}\ F(\th)\ d\th\ ,
\quad n=0,\pm 1,\pm 2,\dots,\quad
\mu_{-n}=\overline {\mu_n} \quad \forall n\ .
\la{trigomoment}
\ea
have a prescribed set of values.} may be formulated as finding the conditions on 
the set of functions,
\ba
\psi(kr) \ = \  \langle e^{-ik\th} \rangle_F\ \equiv\  \frac{1}{2\pi}
\int_0^{2\pi}\!\ e^{-ik\th}
\ F(\th,r)\ d\th\ ,\quad k=0,\pm 1,\pm 2,\dots
\la{fouriercoeffs}
\ea
such that the  function $F(\th,r)$ be positive.  

Let us recall\footnote{We are  here using the polynomial method due to Riesz
\cite{akhiezer,shohat}. On a more general footing, 
it comes from the application 
of
a known theorem
(see \cite{shohat}, Theorem 1.4): 
a necessary and sufficient condition
that the trigonometric moment problem \eqref{trigomoment} have a 
generic ($i.e.$ a solution whose spectrum is not reducible to a finite set 
of points) solution is that all Toeplitz quadratic forms 
\ba
\sum_{j,l=0}^{k}\mu_{j\!-\!l}\ c_j\, \bar c_l >0 , \quad k=0, 1, 2,\dots\ ,
\quad
\forall {\rm \ complex\ vector} \ \vec c
\la{necessary}
\ea
be  positive. This means that the $2k \times 2k$ 
matrices  of moments $\{\mu_{j\!-\!l}\}$ are positive-definite, 
$i.e.$ with  smallest  eigenvalue positive.}  a simple derivation of these 
conditions.
Consider a real positive polynomial  $P(z,\bar z)$, with $z$ being a
complex variable, this polynomial being obtained  
as the squared modulus of an arbitrary complex polynomial $Q(z)$, 
\ba
P(z,\bar z)\equiv \vert Q(z)\vert^2\  = \ \sum_{j,l}^k c_{j}\ z^j \bar z^l\
\bar
c_l\ >\ 0, \quad \forall\ {\rm complex\ vector\ }\vec c\  .
\la{positivepolynoms} 
\ea
Then, choosing $z=e^{-i\th}$ in Eq.\eqref{positivepolynoms} and 
integrating over $F(\th,r)$ as in Eq.\eqref{fouriercoeffs}, one finds,
\ba
 \langle P(e^{-i\th},e^{i\th}) \rangle_F\ =\ \frac{1}{2\pi}\,
\int_0^{2\pi}\!d\th\ \sum_{j,l}^k c_{j}
\  P(e^{i\th},e^{-i\th})\ \bar c_l\  F(\th,r)\ =\  \sum_{j,l}^k\  c_{j}
\ \psi [(j\!-\!l)r]\ \bar c_l\  >\ 0\quad \forall\ \vec c. 
\ .
\la{mupolynoms}
\ea
Then the Toeplitz matrix 
$\{\mathbb M_{jl}\}\equiv\{\psi [(j\!-\!l)r]\}$ associated to 
the quadratic form \eqref{positivepolynoms} with arbitrary coefficients
$c_j$ has to be positive-definite.
More explicitly, for {\it even} functions $\psi$, the Toeplitz matrix
of order $k$, 
\ba
\begin{pmatrix}
\psi(0)& \psi(r)& \psi(2r) & ... & \psi[(k\!-\!1)r]\\
\psi(r)& \psi(0)& ...& ...& \psi[(k\!-\!2)r]\\
...    &     ...& ...&          ...& ...\\
 \psi[(k\!-\!1)r]& \psi[(k\!-\!2)r]& ...& \psi(r)& \psi(0)\\ 
\end{pmatrix}\, ,
\la{Nmatrix}
\ea
is positive-definite, with an eigenvalue spectrum bounded from below by
 zero\footnote{The property \eqref{Nmatrix} appears to be as a necessary 
consequence  to the Bochner theorem
\cite{bochner} applied to the function $\psi(r)$ with a choice  
of points $r_j\!=\!j\cdot r, \ j \in \{1,\dots,k\}\ \forall k \in {\mathbb N}
$ 
Note that thanks to the r-dependence
the condition \eqref{Nmatrix}  on the  Toeplitz matrices  ensures the 
positivity of the Fourier transform, as discussed in the previous section.}.

As an instructive example of conditions resulting from 
the positive-definiteness of the matrices \eqref{Nmatrix}, let us consider 
the case of an even function $\psi(r)$ and its corresponding $3 \times 3$
Toeplitz matrix,
\ba
\begin{pmatrix}
\psi(0)& \psi(r)& \psi(2r)\\
\psi(r)& \psi(0)& \psi(r)\\
\psi(2r)& \psi(r)& \psi(0)\\ 
\end{pmatrix}\ .
\la{3x3}
\ea
Positive-definiteness implies positivity of the matrix determinant
and of its minors along its diagonal. This gives the following set of
inequalities,
\eq
\psi(0)&>&\psi(r)\, ,
\la{minors3}\\
\psi(0)&>&\psi(2r)\ >\ \f{2\psi^2(r)}{\psi(0)}-\psi(0)
\la{major3}\ ,
\eqx
where the last inequality comes from 
the determinant of \eqref{3x3},
\ba
\Dl = [\psi(0)-\psi(2r)]\,[\psi^2(0)-2\psi^2(r)+\psi(2r)\psi(0)]> 0\ .
\la{3determinant3}
\ea
A practical method for positivity tests, coming from the straightforward
generalization to higher order matrices, will be used in the following
sections.

\subsection*{The radial two-dimensional case}
\medskip
Following an approach similar to the one-dimensional case, we want to
relate 
the positivity \eqref{F2d} of the characteristic
function to positive-definiteness properties of sets of 
matrices generalizing (but, actually, not of Toeplitz form) the matrices
\eqref{Nmatrix}.

Let us start with the coefficients of the Fourier series, 
\ba
\psi\left(r\sqrt{m^2+n^2}\right) \ = \  \f {1}{(2\pi)^2}\
\int_0^{2\pi}\!\!\int_0^{2\pi}\!\!\!d\al\, d\be\ e^{-i(m\al+n\be)}\ 
F(\alpha,\beta; r) \ ,
\la{fouriercoeffs2d}
\ea
recasting the characteristic function \eqref{F2d} with appropriate variables.
For this sake, in analogy with the one-dimensional case, we 
consider real positive polynomials built from two complex variables $z,z'$ 
and their complex conjugates, namely 
\ba
P(z,z',\bar z,\bar z')\ \equiv \ \left\vert Q(z,z')\right\vert^2\ =\ 
\left\vert\sum_{j,k}^N c_{j,k} z^j z'^k\right\vert^2 \ =\  
\sum_{j,k,j',k'}\ c_{j,k}\ z^j z'^k \bar z^{j'} 
\bar z'^{k'}\ \bar c_{j',k'}> \ 0\ . 
\la{positivepolynoms2d} 
\ea 
Choosing $(z,z')=(e^{-i\al},e^{-i\be})$ and 
integrating \eqref{positivepolynoms2d} over $F(\alpha,\beta; r)$,  one finds,
\eq 
\langle\, P(e^{-i\al}\!,e^{-i\be}\!,e^{i\al},e^{i\be})\, \rangle_{F}\  &=&
\frac{1}{(2\pi)^2} \ 
\sum_{j,k,j',k'}\ c_{j,k}\ 
\int_0^{2\pi}\!\!\int_0^{2\pi}\!\!\!d\al\, d\be\ e^{-i(j-j')\al-i(k-k')\be}\ 
F(\alpha,\beta; r)\  \bar c_{j',k'}\la{mu}\\ 
&=&\  \ \sum_{j,k,j',k'}\ c_{j,k}\ \psi\left(r\sqrt{(j\!-\!{j'})^2 
+(k\!-\!{k'})^2)}\right)
\  \bar c_{j',k'} >\ 0 \ .
\la{mupolynoms2d}
\eqx 

Formula \eqref{mupolynoms2d} implies the positive-definiteness 
of the tensorial form
\eqref{mupolynoms2d}. A positive-definite matrix form can be obtained 
by noting that  the polynomial $Q(z,z')$ in \eqref{positivepolynoms2d} can
be expanded \cite{lasserre} over the  basis of monomials ordered by their
degree, 
\ba
1\ ,\ z\ ,\ z'\ ,\ z^2\ ,\ zz'\ ,\ z'^2\ ,\ z^3\ ,\ z^2z'\ ,\ zz'^2\ ,z'^3\ ,
\dots
\la{basis}
\ea

The positivity condition \eqref{mupolynoms2d} reads
as a positive, quadratic\footnote{It is interesting to note 
that polynomials of two variables are not necessarily sums of 
squares  but can always be expressed as a ratio of sum of 
squares \cite{lasserre}. Hence it is enough to ask for an
arbitrary squared polynomial 
\eqref{positivepolynoms2d}.} form and thus as the definite-positiveness of an
 ordered hierarchy of matrices whose sizes depend on the chosen maximal 
orders of the corresponding polynomial \eqref{positivepolynoms2d}.

Let us illustrate this property by a low degree case, namely the 
positive definiteness of the $3\times3$ 
matrix,
\ba
\begin{pmatrix}
\psi(0)& \psi(r)& \psi(r\sqrt 2)\\
\psi(r)& \psi(0)& \psi(r)\\
 \psi(r\sqrt 2)& \psi(r)&\psi(0)\\ 
\end{pmatrix}\ .
\la{3matrix}
\ea
Positive-definiteness implies positivity of the matrix determinant
and of its minors along its diagonal, hence,
\eq
\psi(0)& > &\psi(r)\, ,
\la{minors}\\
\psi(0)& > &\psi(r\sqrt 2)\  > \ \f{2\psi^2(r)}{\psi(0)}-\psi(0)
\la{major}\ ,
\eqx
where the last inequality comes from the
 determinant of \eqref{3matrix}
\ba
\Dl = [\psi(0)-\psi(r\sqrt 2)]\, [\psi(r\sqrt 2)\psi(0)-2\psi^2(r)+
\psi^2(0)] 
>  0\ .
\la{3determinant}
\ea
Comparing with the similar one-dimensional case \eqref{3determinant3}, it
is worth noting that the minors' inequalities from the matrix minors are
the same as the one-dimensional ones \eqref{mu}, up to a rescaling of r.
It is not the case for the last inequality of  \eqref{major}, coming from the
 determinant \eqref{3determinant}. Indeed, differences obviously occur 
because the new matrices, starting with \eqref{3matrix} and beyond, are not
 any more of 
a Toeplitz type. Such differences will be the common rule at higher orders.

The structure of the set of inequalities (\ref{minors},\ref{major})
can be elucidated by remarking that it stems from the 
application of the Bochner theorem to the set of positions,
\ba
\vec {x_j} \ =\ \{0,0\}\ ,\ \{0,r\}\ ,\ \{r,0\},
\la{points}
\ea
in a 2-dimensional square lattice. Indeed, using the
2-dimensional  Bochner theorem \cite{bochner}, one finds 
the related necessary condition of positive-definiteness 
on the matrices (here a $3\times3$ matrix) of 2-vectors
\ba
\{M_{j,l}\}\ \equiv\  \{\, \psi(\vec{x_j}\!-\!\vec{x_l})\,\}\ . 
\la{2dbochner}
\ea
It is easy to recognize the identity of \eqref{2dbochner} 
with the matrix \eqref{3matrix}. Moreover typical minor's 
inequalities  correspond to the Bochner theorem
for the pairs of points $(\{0,0\} , \{1,0\})$ 
and $(\{0,0\} , \{1,1\}),$ respectively. This explains their relation with
the one-dimensional properties. Putting together the three points  $\{0,0\} ,
 \{1,0\} , \{1,1\},$ which are not aligned, gives rise to the matrix 
\eqref{3matrix}
which is not of Toeplitz form, characteristic of the one-dimensional problem.

The same arguments easily extend to higher orders. For 
instance, at the next level,  $d=2$, one comes to a $6\times6$ 
positive-definite matrix of 2-vectors
 $\{ \psi(\vec {x_j}\!-\!\vec{x_l})\}$ corresponding to the basis
\eqref{basis} with six 2-vectors
\ba
\vec{x_j} \ =\ \{0,0\}\ ,\ \{0,r\}\ ,\ \{r,0\}\ ,\ \{0,2r\}\ ,\ \{r,r\}\ ,
\ \{2r,0\}\ .
\la{6points}
\ea
As in the one-dimensional case, the generalization to higher degrees is 
relatively straightforward and will lead to a subsequent 
application to Fourier-positivity in the radial two-dimensional case.

\section{Applications to Fourier-positivity}
\la{applis}

As theoretically motivated and developed in the preceding sections, 
Fourier-positivity of a real and even, positive function $\psi(r)$ can be
tried and 
checked using in  a finite set of rescalings of $\psi,$ namely
$\mathbb S \equiv \{\psi(kr)\},\ k=1,\dots,K.$ In 
a first way that we call in short ``Bochner method'', we make use of 
the positive-definiteness of $r$-dependent matrices whose components 
are given by  $\mathbb S$, 
as discussed in  section \ref{poissonbochner}, see \eqref{Nmatrix} for
 the one-dimensional case  and \eqref{2dbochner}
for the radial two-dimensional case.

The second way to test Fourier-positivity using the tools of section 
\ref{poissonbochner} makes a direct use of the  characteristic functions 
stemming from the Poisson resummation formulas, see respectively \eqref{sum} 
for one-dimensional cases and 
\eqref{iff2d2}
for the radial two-dimensional cases. The idea is to look for the appropriate
range of the variable
$r$, see resp. \eqref{range1} and \eqref{range2}, for which the
reconstruction 
of the characteristic functions from 
a finite set $\mathbb S$ satisfies the 
selection of the Fourier transform $\va(s)$ with sufficient 
accuracy to detect possible violations of positivity in some range of $s.$

We shall test the capacity of the two 
different methods to thoroughly check
Fourier-positivity or its violation. For this sake we introduce a
large testing set  of {\it real}, {\it even} and {\it positive}
functions $\psi(r),$ Fourier-positive or not, made of random combinations
of a finite basis with
well-known analytic Fourier transforms. To be more precise, we are using, 
in the one-dimensional case,
the orthogonal basis of Hermite-Fourier functions,
i.e. the quantum oscillator 
eigenstates which are eigenstates of the Fourier transform
with eigenvalues 1 and -1. In the radial two-dimensional case we opt for
an orthogonal  basis of Laguerre polynomials 
multiplied by a simple exponential; these, under Fourier-Bessel transform,
return combinations of rational functions fast decreasing at infinity.
The random
nature of the chosen combinations allow us to start with a large corpus  of
Fourier-positive and non Fourier-positive functions $\psi$, allowing us
to test our methods
with a good accuracy and for quite different sets of Fourier partners.

\subsection{Fourier-Positivity in one dimension}

Starting with a large set of real even test functions, we examine the
performance 
of the Fourier-positivity tests 
corresponding successively to the ``Bochner method'' and the ``Poisson
method''.

\subsubsection*{A one-dimensional basis of Hermite-Fourier test functions}

Consider the Hermite-Fourier functions
\ba 
u_p(r)=\pi^{-\frac{1}{4}}\, e^{-\frac{1}{2}r^2} H_p(r).
\la{hermite0}
\ea
Here, we set $H_p$ to be a {\it square normalized} Hermite polynomial,
with a 
positive coefficient for its highest power term. For the sake of clarity, we 
list the first polynomials as, $H_0=1,\, H_1=\sqrt{2}\ r,\, 
H_2=(2r^2-1)/\sqrt{2},\, H_3=(2r^3-3r)/\sqrt{3}\, $, and their recursion 
relation,
\ba
a_{p+1}H_{p+1}=2\, r\,  a_{p}H_{p} -2\, n\, a_{p-1}H_{p-1}\ ,
\la{Hermite}
\ea
where $a_{p}= \sqrt{2^p p!}\ .$ It is known that the Fourier transform of 
such states brings 
only a phase,
\begin{equation}
\frac{1}{\sqrt{2\pi}}\, \int_{-\infty}^{\infty} dr\ e^{i s r}\, u_p(r) = 
i^p\ u_p(s)\, ,
\label{transpa1}
\end{equation}
and thus such states give generalized self-dual functions with phase $i^p.$ 
If one expands $\psi$ in the oscillator basis, 
$\psi(r)=\sum_{p=0}^N c_p\, u_p(r),$ with a truncation at some degree $N,$ 
then all odd order components $c_{2p+1}$ must vanish if $\varphi$ must
be real, 
and the even rest splits, under Fourier transform, 
into an invariant part and a part with 
its sign reversed, namely
\eq
\psi(r) &=& \sum_{p=0}^{[N/4]} \ c_{4p}\  u_{4p}(r)\ +\  
\sum_{p=1}^{[N/4]} c_{4p-2}\ u_{4p-2}(r),\n
\varphi(s) &=& \sum_{p=0}^{[N/4]} \ c_{4p}\  u_{4p}(s)\ -\  
\sum_{p=1}^{[N/4]} c_{4p-2}\ u_{4p-2}(s)\, ,
\label{transfo1}
\eqx
where the usual symbol $[N/4]$ means the 
integer part of $N/4$. This polynomial parametrization makes it trivial
to generate fully positive $\psi$'s, with both cases of partners $\va$'s
fully positive or $\varphi$'s showing both signs.

In our numerical illustration
we consider the basis with $N=8,$ i.e. random combinations of the first
five real eigenstates of the 
harmonic oscillator\footnote{Our explicit parametrization is, 
\begin{eqnarray}
\psi(r)\ =\ \pi^{-\frac{1}{4}}
e^{-\frac{1}{2} r^2}\, [\, c_0 + 
c_2\,  2^{-\frac{1}{2}} (2r^2-1) + 
c_4\,  6^{-\frac{1}{2}} (4r^4-12r^2+3)/2 +  \nonumber \\ 
c_6\,  5^{-\frac{1}{2}} (8r^6-60r^4+90r^2-15)/12 +
c_8\, 70^{-\frac{1}{2}} (16r^8-224r^6+840r^4-840r^2+105)/24\, ]\, ,
\la{psip}
\end{eqnarray} and
with Fourier transform,
\begin{eqnarray}
\varphi(s)\ =\ \pi^{-\frac{1}{4}}
e^{-\frac{1}{2} s^2}\, [\, c_0 - 
c_2\,  2^{-\frac{1}{2}} (2s^2-1) + 
c_4\,  6^{-\frac{1}{2}} (4s^4-12s^2+3)/2 - \nonumber \\
c_6\,  5^{-\frac{1}{2}} (8s^6-60s^4+90s^2-15)/12 +
c_8\, 70^{-\frac{1}{2}} (16s^8-224s^6+840s^4-840s^2+105)/24\, ]\, .
\la{phip}
\end{eqnarray}
} whose
coefficients, $\{c_0,c_2,c_4,c_6,c_8\}$, are random real numbers with a
normalization constraint, $c_0^2+c_2^2+c_4^2+c_6^2+c_8^2=1$, and the
condition
that $\psi(r) > 0,\, \forall r$. We retained a randomly generated set 
of 15456 positive functions $\psi(r)$ among which  4388 cases where
$\varphi$ is
also always positive and 11068 cases where $\varphi$ has a range of 
negative values.

\begin{figure}
\scalebox{.65}{\includegraphics{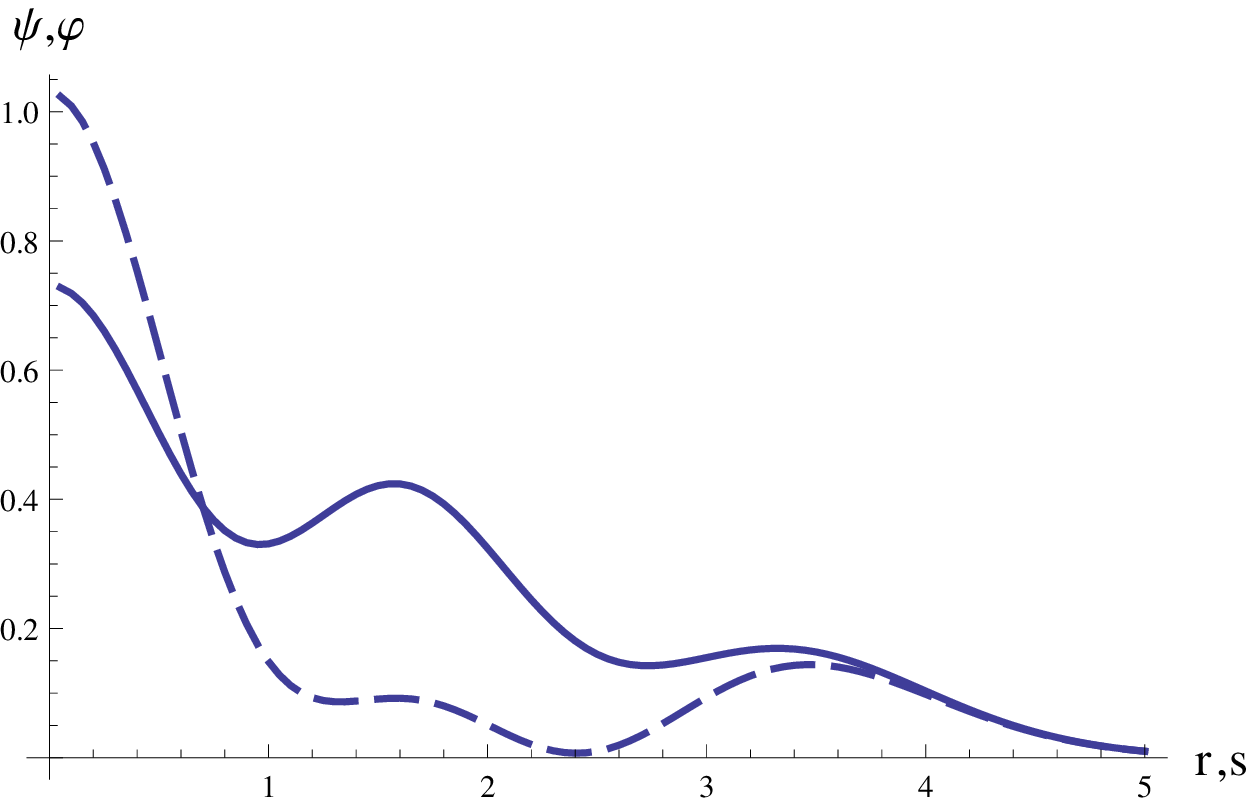}}
\scalebox{.65}{\includegraphics{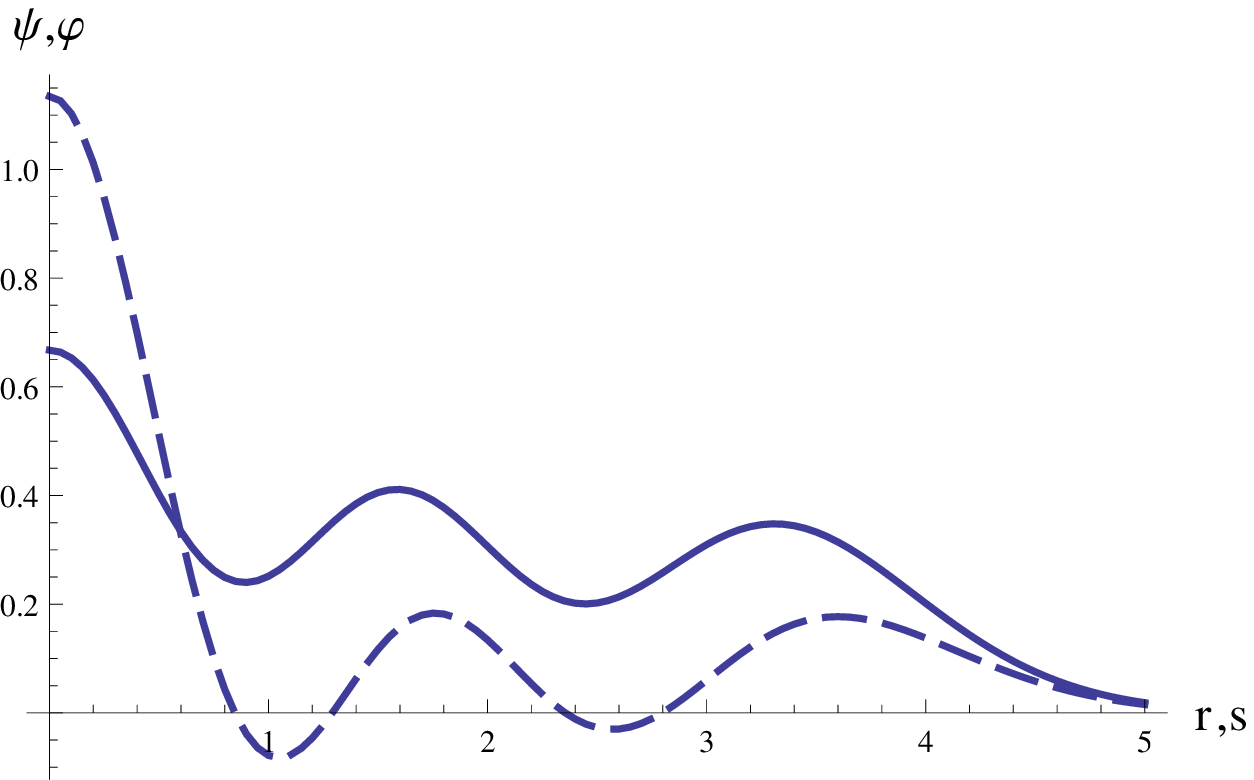}}
\caption{{\it Examples of  ``positive-positive'' and 
``positive-negative''  test functions.} Left: both $\psi_{pp}$ (full line),
with components,
$\{.901,.276,.259,.006,.214\}$, and $\varphi$ (dashed line) 
are positive. Right: $\psi_{pn}$, with components, 
$\{.772,.304,.386,.171,.366\}$, 
remains $>0$ but $\varphi$ takes both signs.} 
\la{1}
\end{figure}

\subsubsection*{One-dimensional Bochner method}

For further discussions, let us denote  $\psi_{pp}(r),$ the
``positive-positive''  (resp.
``positive-negative'' $\psi_{pn}$)  test functions which lead to
$\varphi$  positive (resp. not everywhere positive).
Figure \ref{1} shows a typical example in each category.

\begin{figure}
\scalebox{.65}{\includegraphics{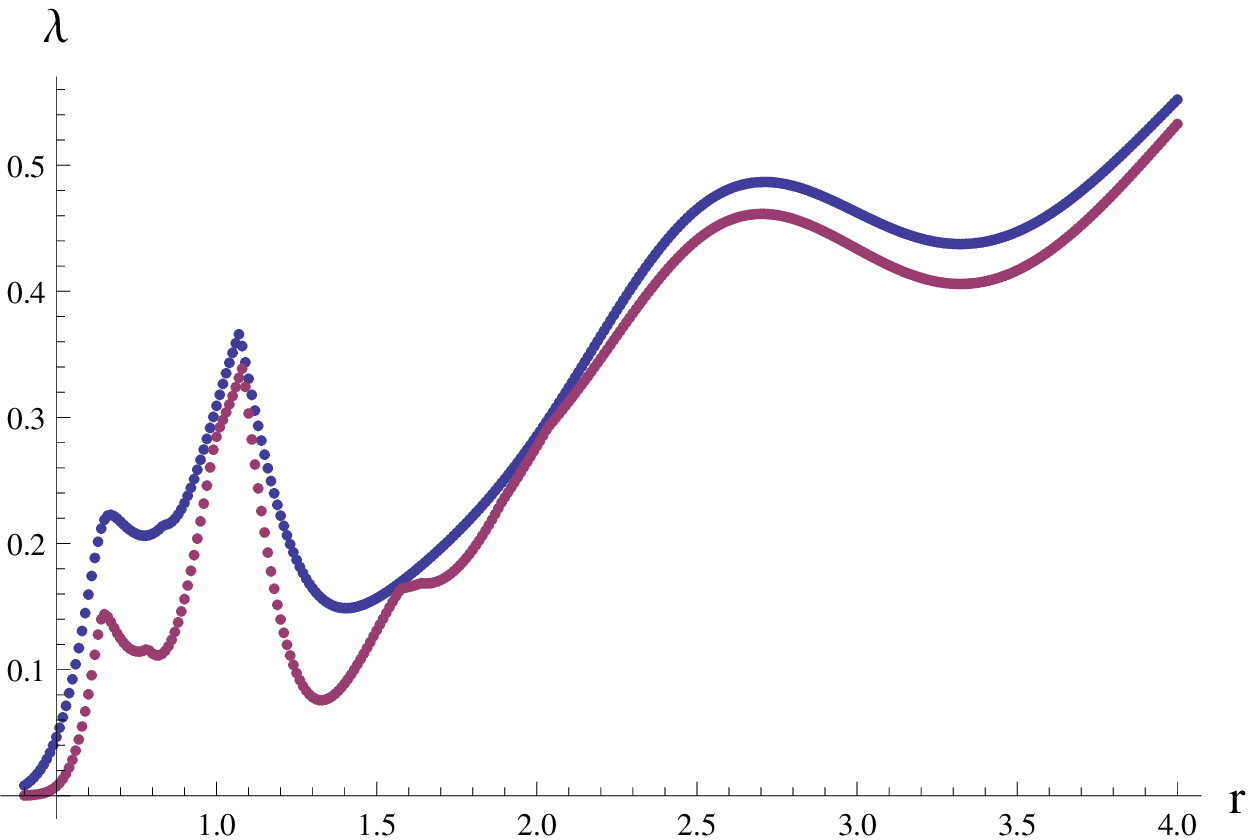}}
\scalebox{.65}{\includegraphics{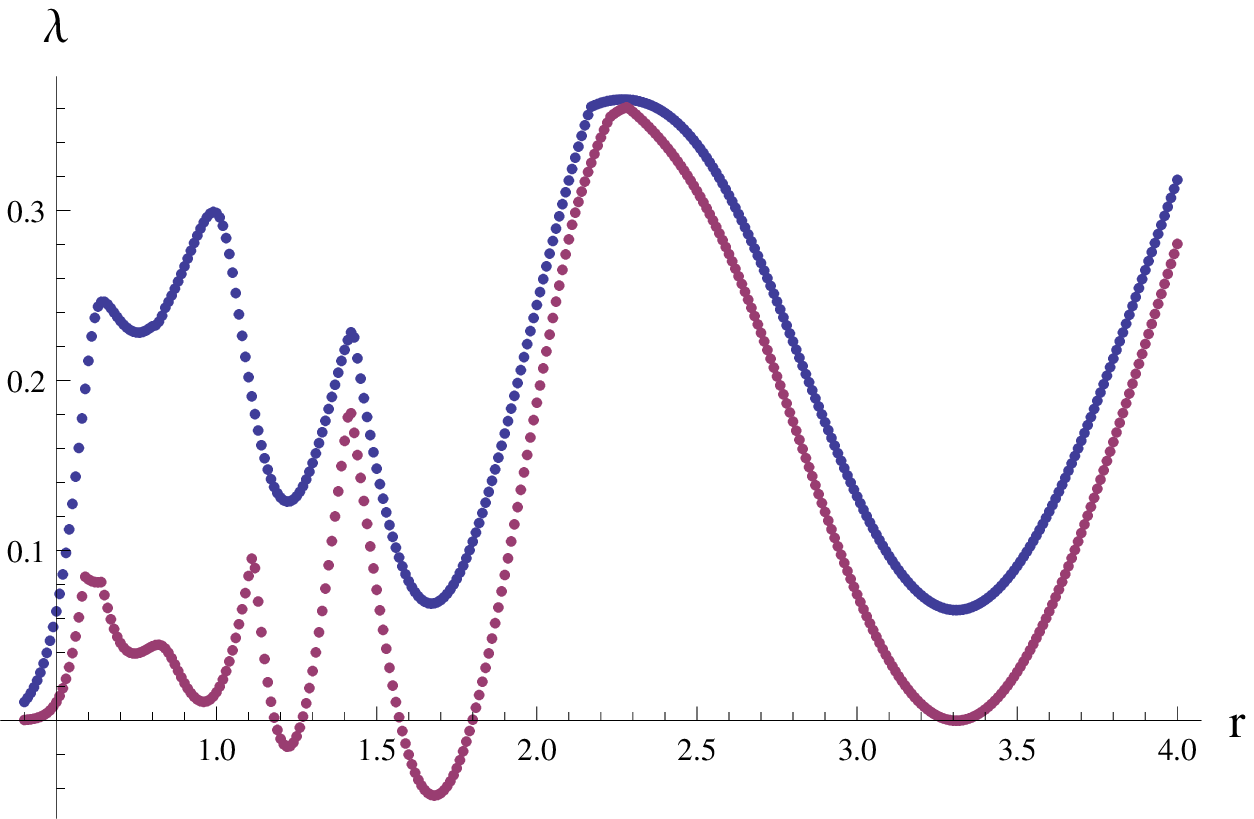}}
\caption{{\it Lowest eigenvalue $\lambda$ as a function of the ``grid''
parameter 
$r$ of the Toeplitz matrices, see \eqref{Nmatrix}}. Upper curves, matrix
 dimension $5$. Lower
curves,
matrix dimension $10$. Left: $\psi_{pp}$ case, as described by left part of
Fig.\ref{1}. Right: $\psi_{pn}$ case shown in right part of Fig.\ref{1}, 
non positivity of $\varphi$
undetected by dimension $5$, detected by dimension $10$.}
\la{2}
\end{figure}

Following the results of section \ref{poissonbochner}, all Toeplitz
matrices \eqref{Nmatrix} built from  $\psi_{pp}(r)$ are
positive definite, for all dimensions and for all values of the parameter
$r$.
None of our numerical tests based on positive-definiteness with our sample
 of 4388 functions $\psi_{pp}(r)$
contradicted this fact.

According to the same properties, the lack 
of positive definiteness  
of the Toeplitz matrix corresponding to the functions $\psi_{pn}(r)$
will show off, sooner or later, 
for some dimension and some
range of $r$. The sign of the Toeplitz determinant, however, may be
misleading if an even number of negative eigenvalues occurs\footnote{A way
to evade this difficulty would be to increase the dimension by one,
but this costing computation time.}. It is
safer to track the lowest eigenvalue of a high enough 
dimensional matrix of the hierarchy \eqref{Nmatrix}. 

 Figure \ref{2}
shows, for dimensions $5$ and $10$, how this lowest eigenvalue, 
$\lambda_5,\lambda_{10}$, respectively, behaves when $r$ varies when choosing the example functions of Fig.\ref{1}. Obviously,
$\lambda_{10} \le \lambda_5,$ since the 5-dimensional Toeplitz matrix 
is embedded in any higher-dimensional matrix of the 
hierarchy \eqref{Nmatrix}. The left part of Fig.\ref{2} considers again that
case
that was
described by the left part of Fig.\ref{1}. As should be, no negativity
is observed, and a confirmation is expected for any matrix dimension.
In turn, the right part of Fig.\ref{2} describes that case
that was illustrated by the right part of
Fig.\ref{1}. No detection occurs if the Toeplitz matrix has only
dimension $5$, while dimension $10$ provides a clear detection. 

Across our sample of $11068$ cases, a proportion of $\sim 69\%$ are detected
with Toeplitz matrices of dimension $5$. With dimension $10$, the success
rate reaches $\sim 86\%$. The deeper the negative parts of $\varphi$, the
higher the detection probability. However, the ``Bochner method'' 
may require for some ``rebel'' functions a very high dimensionality, 
and then becomes uneasy. The ``Poisson method'' will turn out
to be easier and with
almost full detection success.

\subsubsection*{One-dimensional Poisson method}

Consider the parameter $r$ as an integration grid parameter, $\Delta r$,
and also the ratio, $\theta/\Delta r$, as a pseudo momentum, $s$.
Using the renormalized definition of the characteristic
function \eqref{sum} and the approximation \eqref{approxfourier1}
 discussed in subsection \ref{oned}, the range of $\psi$ in our samples
allows a truncation into a finite sum, namely,
\ba
F(s,\Delta r)\ \simeq\ \Delta r/\sqrt{2 \pi}\, 
\sum_{n=-K}^K 
\psi(n \Delta r)\ \exp[i (n \Delta r) s]
\la{renorm}
\ea
with $K \simeq R/(\Delta r)$. Here a
range $R=10$ is enough to perform a correct coverage of the 
characteristic function $F(s,\Delta r).$ Note that we thus keep $K\Delta r
\simeq R$ constant
and look for an integration grid parameter $\Delta r$ not too small, 
bounded from below, as discussed in subsection \ref{oned}. 

Contour lines of the values of $F$, in terms of $\Delta r$ and $s$, are
shown
 in Figure \ref{3}. Its left and right parts correspond to the 
``doubly positive''
and ``partly negative'' cases already used for the previous Figures. As 
expected, no contour for a negative value of $F$ is found for the ``doubly
positive'' case, while ``negative contours'' occur for the case when
$\varphi$
is partly negative. Note, for instance, the
``negative'' contour line where $F=-0.028$.  A few big dots
have been plotted to reinforce the visual identification of such
``negative'' contours.

Such a result, namely a good approximation to a brute force Fourier transform,
is expected when $\Delta r$ is small enough
to ensure a good convergence, $F(s,\Delta r)\rightarrow \varphi(s)$, but
larger
values of $\Delta r$ maintain the criterion: negative values of $F$ occur
only for non positive $\varphi$'s.

A comment is in thus order: It can be remarked from Fig.\ref{3} that the
contour 
curves remain parallel to the abscissa axis for a rather large range of
$\Del r.$ 
This, joined to the previous remark, can be explained by the fact that the
 Fourier transform $\va(s)$ can be quite well reconstructed from the
approximation 
\eqref{renorm} independently from the value of $\Del r$ or equivalently of
the 
summation number $K.$   This is exemplified in Figure \ref{4} for the
non Fourier-positive 
function of Fig.1 (right). This is why negative values are 
reproduced and the test for non Fourier-positivity is successful.

We verified, for $\Delta r$ running between $.1$ and $1$ and $0 < s < 8$,
that our $4388$ ``doubly positive'' test cases did not generate any negative
value of $F$. In turn we verified, for $\Delta r$ running between $.1$ and
$1$
and $0 < s < 8$ again, that most partly negative $\varphi$'s are detected
by negative contours. Typically, for $11068$ test cases, only $19$ of them
fail generating negative values of $F$ and thus escape detection. An 
inspection of such ``rebel'' cases gives a clear explanation for the failure:
the negative values of such $\varphi$'s are tiny.

It can be concluded that $F$ provides a very efficient test for the
selection of $\psi$ without a detailed calculation of $\varphi$.

\begin{figure}
\scalebox{.65}{\includegraphics{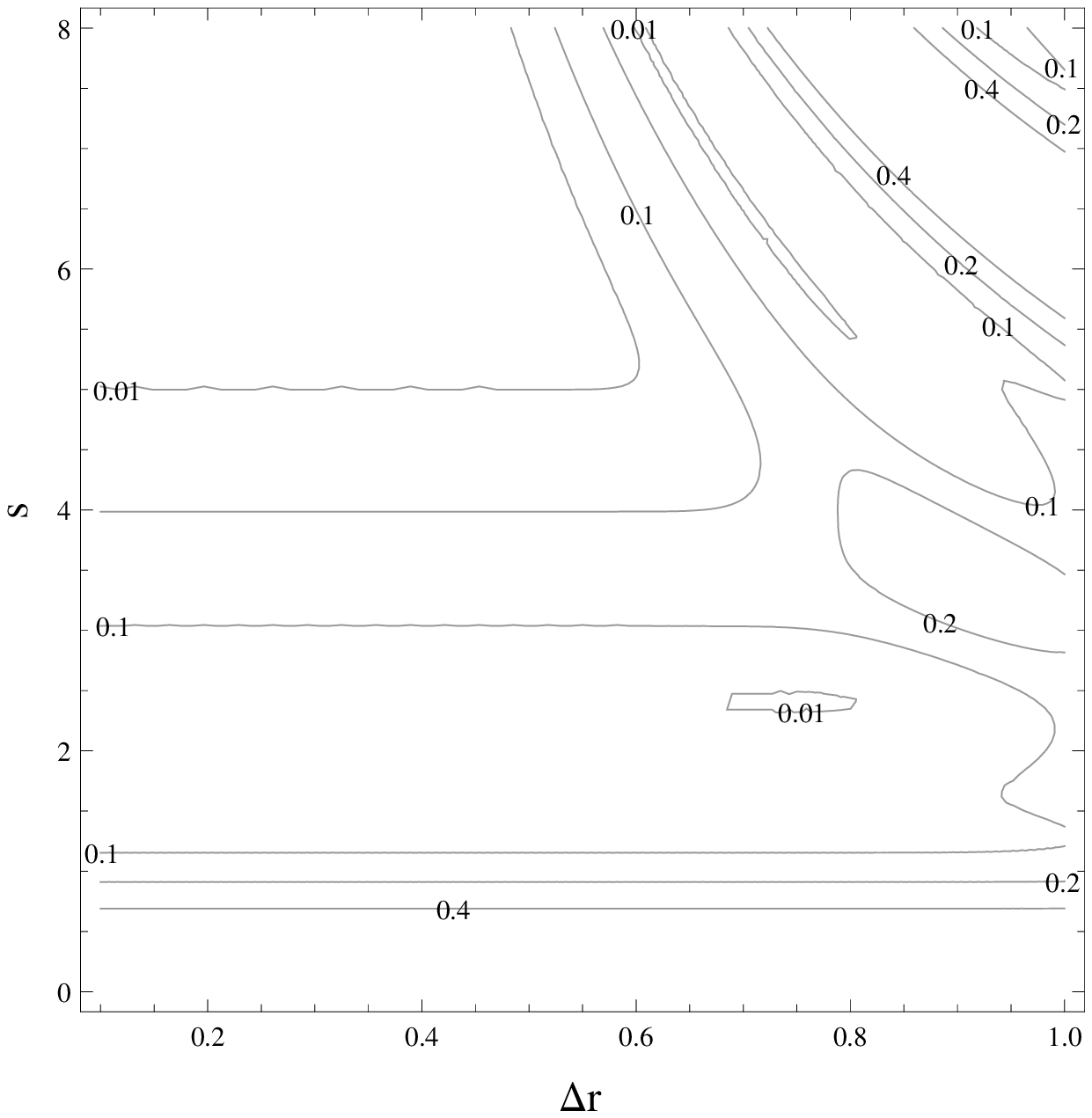}}
\scalebox{.65}{\includegraphics{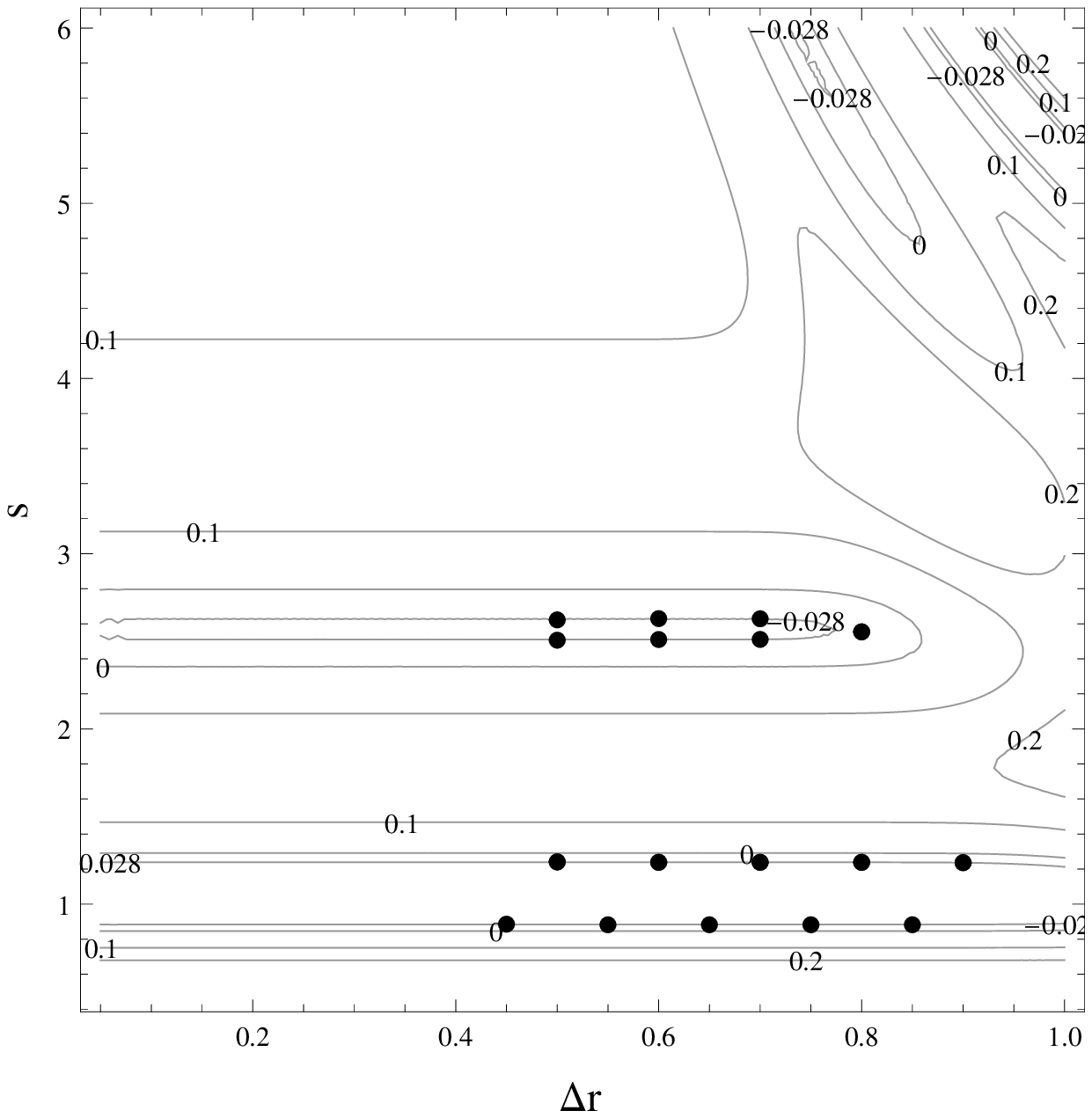}}
\caption{{\it Contour plots of the characteristic function $F$ in terms of
the
discretization parameter $\Delta r$ and the pseudo-momentum 
$s \equiv \theta/\Delta r$.} Left: Double positivity case, already 
described by left part of Fig.1. No contour line is found for negative 
values of $F$. Right: Case already shown in right part of Fig.1, 
non positivity of $\varphi$ detected by contours for negative values
of $F$.}
\la{3}
\end{figure}

\begin{figure}
\scalebox{.65}{\includegraphics{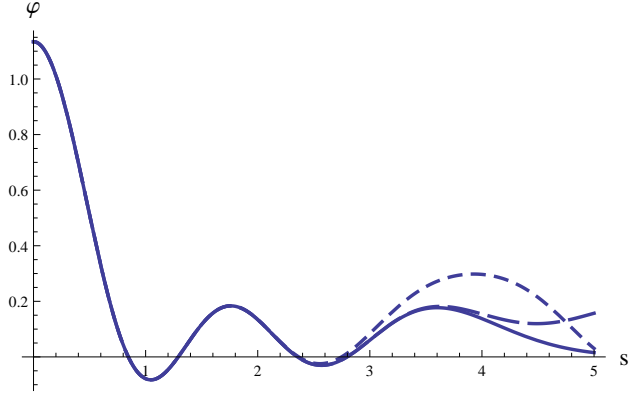}}
\caption{{\it Example of a `reconstruction'' of a
non-Fourier-positive test function.} The function $\varphi_{pn}$, depicted 
in the right part of Fig.1 by a dashed line,
is approximated respectively
with  $K=12$ (small dashes), $K=14$ (large dashes) and $K=20$ (full line,
indistinguishable from the true $\varphi_{pn}$.).}  
\la{4}
\end{figure}

\subsection{Fourier-Bessel-positivity} 
\subsubsection*{Basis of radial functions connected by Bessel transform}

Using a similar method as in one dimension, we set a basis, for the
functions
$\psi$, built from  the exponential, $e^{-\frac{x}{2}}$, multiplied 
by random linear combinations of nine Laguerre
polynomials\footnote{The explicit form of the basis is \\
$
e^{-\frac{x}{2}}\times\{1,\, (-2 + x)/\sqrt{2},\, (6 - 6 x + x^2)/(2\sqrt{3}),
\, 
 (-24 + 36 x - 12 x^2 + x^3)/12,\, (120 - 240 x + 120 x^2 - 20 x^3 + x^4)/
  (24 \sqrt{5}),\, (-720 + 1800 x - 1200 x^2 + 300 x^3 - 30 x^4 + x^5)/
  (120 \sqrt{6}),\, (5040 - 15120 x + 12600 x^2 - 4200 x^3 + 630 x^4 - 42 x^5
 + x^6)/(720 \sqrt{7}),\, (-40320 + 141120 x - 141120 x^2 + 58800 x^3 - 
   11760 x^4 + 1176 x^5 - 56 x^6 + x^7)/(10080 \sqrt{2}),\, 
 (362880 - 1451520 x + 1693440 x^2 - 846720 x^3 + 211680 x^4 - 28224 x^5 + 
   2016 x^6 - 72 x^7 + x^8)/120960)\}
$.}. These are normalized,
$\int_0^{\infty} x \psi(x)^2 dx=1$, from the condition, $\sum_i c_i^2=1,$
for 
the random, real number, mixture coefficients, $c_i,i=0,\dots,8$. The same
coefficients are also selected so that $\psi(x)$ be positive,
$\forall x\ge 0.$
The partners in radial momentum space, 
$\varphi(p)=\int_0^{\infty}  x J_0(p x) \psi(x) dx$, are obtained, with the
same coefficients $c_i$, from the corresponding basis of
functions\footnote{The basis in Fourier transformed space reads :\\
$
\{4/(1 + 4 p^2)^{3/2},\, (-4 \sqrt{2} (-1 + 8 p^2))/(1 + 4 p^2)^{5/2},\, 
 (4 \sqrt{3} (1 - 24 p^2 + 48 p^4))/(1 + 4 p^2)^{7/2},\, 
 (-8 (-1 + 48 p^2 - 288 p^4 + 256 p^6))/(1 + 4 p^2)^{9/2},\, 
(4 \sqrt{5} (1 - 80 p^2 + 960 p^4 - 2560 p^6 + 1280 p^8))/(1 + 4 p^2)^{11/2},
\, 
(-4 \sqrt{6} (-1 + 120 p^2 - 2400 p^4 + 12800 p^6 - 19200 p^8 +
6144 p^{10}))/
  (1 + 4 p^2)^{13/2},\, (4 \sqrt{7} (1 - 168 p^2 + 5040 p^4 - 44800 p^6 + 
    134400 p^8 - 129024 p^{10} + 28672 p^{12}))/(1 + 4 p^2)^{15/2},\, 
 (-8 \sqrt{2} (-1 + 224 p^2 - 9408 p^4 + 125440 p^6 - 627200 p^8 + 
    1204224 p^{10} - 802816 p^{12} + 131072 p^{14}))/(1 + 4 p^2)^{17/2},\, 
 (12 (1 - 288 p^2 + 16128 p^4 - 301056 p^6 + 2257920 p^8 - 7225344 p^{10} + 
    9633792 p^{12} - 4718592 p^{14} + 589824 p^{16}))/(1 + 4 p^2)^{19/2}\}
$.}
, also normalized.
It is clear that $\varphi(p)$ reads as a polynomial divided by a common
denominator, $(1 + 4 p^2)^{19/2}$. This makes it easy to sort out positive
$\varphi$'s from those which take both negative and positive values. We show
in Figure \ref{5} a double positivity case (left part) 
and a case with $\varphi$ partly negative (right part). 

\begin{figure}[b]
\scalebox{.65}{\includegraphics{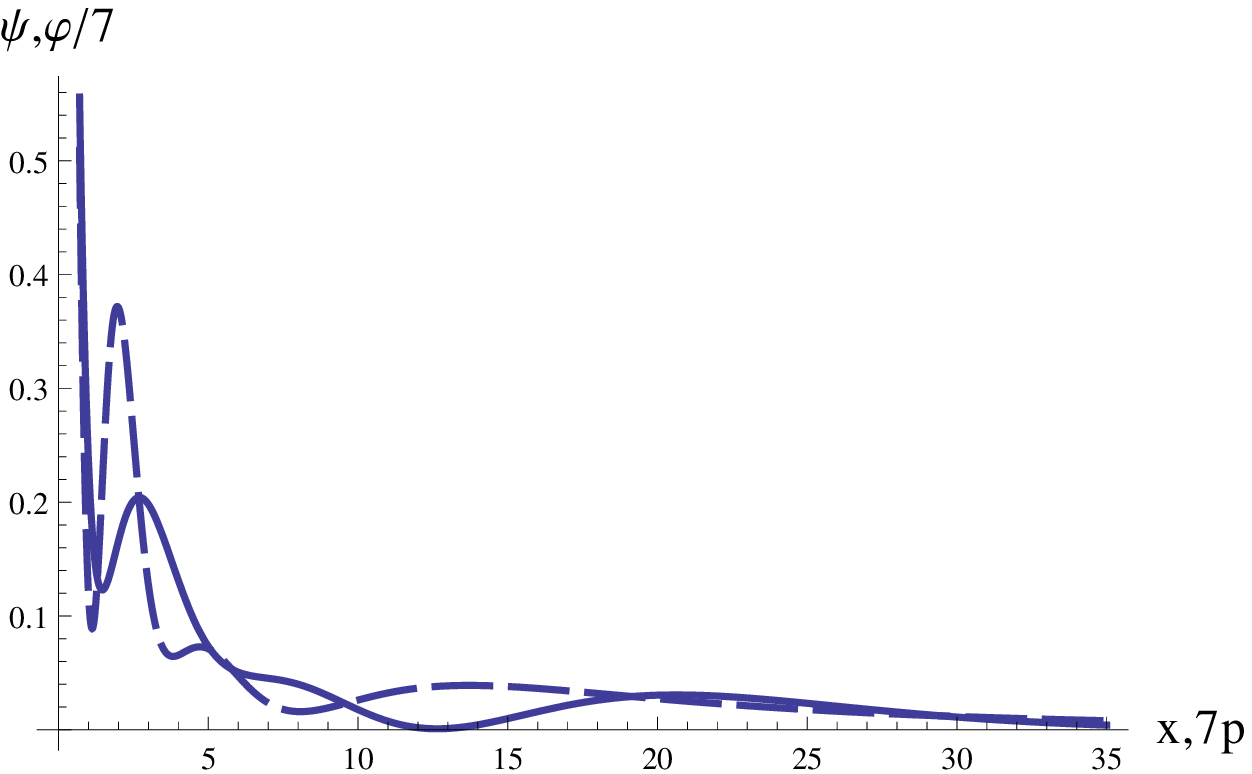}}
\scalebox{.65}{\includegraphics{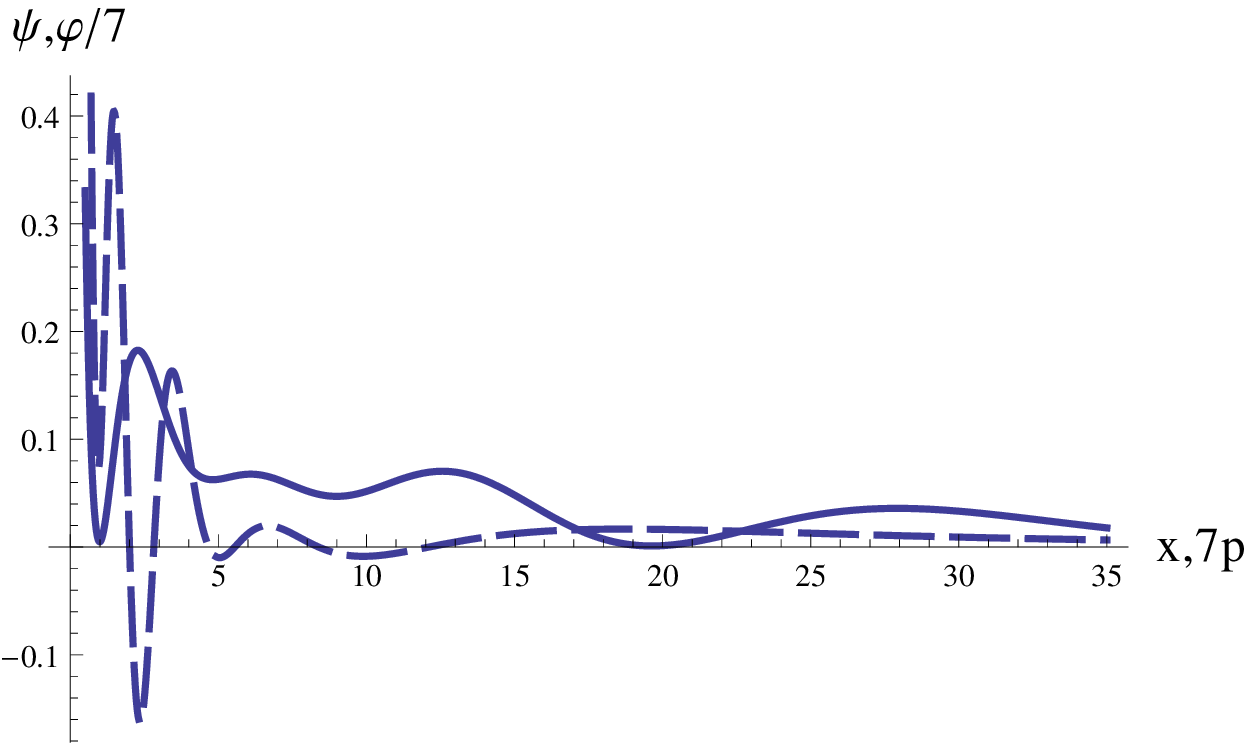}}
\caption{{\it Radial partners under Bessel transform.} 
$\psi$ (full lines) and 
$\varphi$ (dashes). 
$Left:$ Fourier-positive case 
$\psi=e^{-\frac{x}{2}}(3.6096 - 6.7462 x + 4.8826 x^2 - 1.6141 x^3 +
0.28086 x^4 -  0.027009 x^5 + 0.00145 x^6 - 0.000042169 x^7 + 
5.63539\, 10^{-7} x^8).$
$Right:$ non Fourier-positive case 
$\psi=e^{-\frac{x}{2}}(2.49362 - 6.84573 x + 6.76697 x^2 - 3.04127 x^3 +
0.723816 x^4 - 0.0959944 x^5 + 0.00705616 x^6 - 0.000265057 x^7 +
3.93896\,10^{-6} x^8)$. 
For graphical reasons, the Figure actually shows 
$\varphi(p/7)/7$.}
\la{5}
\end{figure}

Our test basis for double positivity contains $185$ cases and that for
situations where only $\psi$ remains always positive contains $9894$ cases.

\subsubsection*{Bochner method for radial functions}

It must be kept in mind here that we are in a two-dimensional situation,
namely
that, given a point with coordinates $\{y,z\}$, the argument $x$ of $\psi$
is, $x=\sqrt{y^2+z^2}$. Given a list of points $\{y_i,z_i\}, i=1,\dots,K$,
the matrix elements of the associated, $K$th-order Bochner matrix read,
$\psi\left( \sqrt{(y_i\!-\!y_j)^2+(z_i\!-\!z_j)^2} \right)$. Our numerical
tests
used a set of points $\vec r_i$ the coordinates of which, $\{y_i,z_i\}$, are
random real numbers in a range, $\{-20,20\}$. A scale parameter, $\beta$,
is then introduced to adjust such points to any range 
$\{-20 \beta,20 \beta\}$.
Typically, we considered the first 20 points, $\{\beta y_i,\beta z_i\}$,
with
for instance $\beta=.5$, but we also used $\beta=.1$, $\beta=.4$,
$\beta=.8$,
$\beta=1$. When 20 points gave too few detections, we used as many as $80$
or
even $100$ points. 

As expected, the Bochner matrices are found positive definite when we 
investigate the $185$ ``doubly positive'' cases. In turn, our set of $9884$
test functions for partly negative $\varphi$'s returned a detection rate of
$\sim 20\%$ when Bochner matrices of dimension $20$ were used. The rate
reached
$\sim 40\%$ for matrices of dimension $80$ and hardly increased if $100$
points were used.  This modest detection rate with reasonable size matrices
contrasts with the good result obtained in the one-dimensional, Gaussian
test function case. Matrices of a much higher order are
now needed, or a set of more efficient points $\vec {r_i}$ must be defined.
But we
tried several sets of points, different from random ones, and failed to
design a ``maximum efficiency set of points''.

Fig. \ref{6} shows, for dimensions $20,40,60,80$ respectively, the evolution
of
the lowest eigenvalue $\lambda$ of a Bochner matrix as a a function of the
scale parameter $\beta$. Naturally, the matrix with dimension $20$ being a
submatrix of that with dimension $40$, the latter generates a lower bound. 
The same reasoning applies when the dimension increases, hence the ordering
of the four shown curves. The left part of the Figure describes a case where
negative values of $\lambda$ are fast obtained. The left part describes a
failure case (even for dimension 100).
\begin{figure}
\scalebox{.65}{\includegraphics{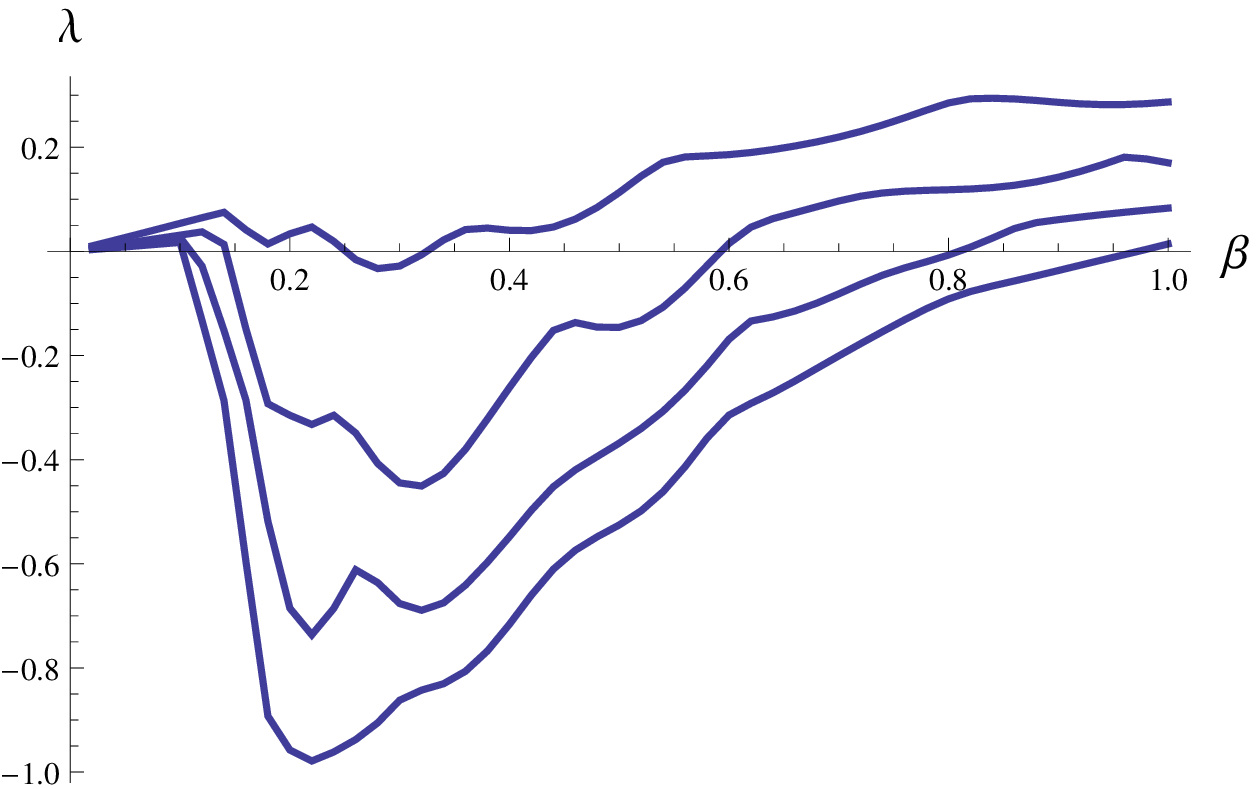}}
\scalebox{.65}{\includegraphics{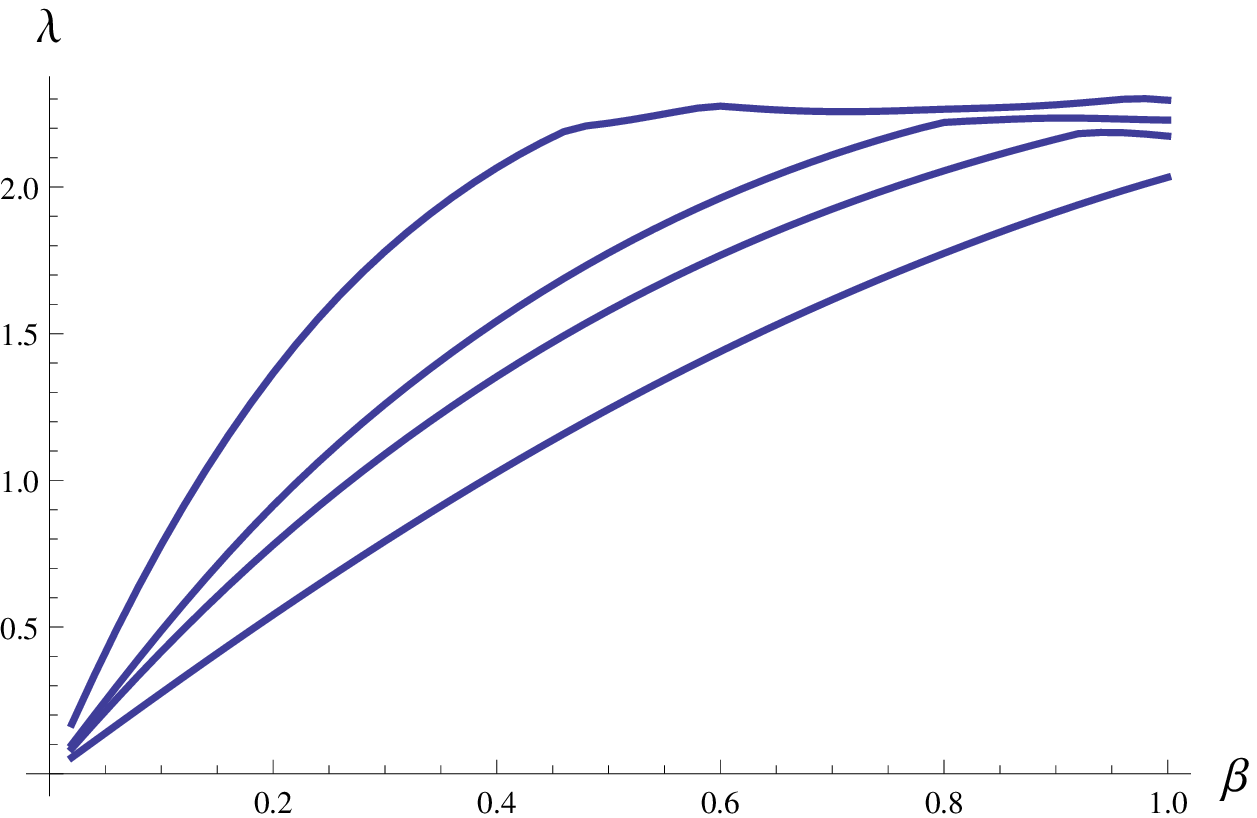}}
\caption{{\it Evolution of the lowest eigenvalue of a Bochner matrix as a
function
of the scaling parameter $\beta$ for the random points which parametrize the
matrix.} From top to bottom curve, dimension $20,40,60,80$. Left: the
eigenvalue soon becomes negative and detects a partly negative $\varphi$. 
Right: failure case, where a much larger dimension would be needed for the
detection.}
\la{6}
\end{figure}

\subsubsection*{Poisson method for radial functions}

With the obvious symmetries at our disposal, the Poisson function in this
situation can be rewritten as,
\begin{equation}
F(\alpha,\gamma,\Del r)=(\Delta r)^2/(2 \pi) \sum_{m=0}^K \sum_{k=0}^K  
e_m e_n
\psi\left(\Delta r \sqrt{m^2+n^2}\right) \cos (m \alpha) \cos (n \gamma)\, ,
\la{eqradial}
\end{equation}
where, $e_m=2-\delta_{m0}$, and the same for $e_n$\! , account for edge
effects.
The range $R$ of $\psi$, typically $R \simeq 40$ for our test functions,
provides a natural cut-off, $K \simeq R/\Delta r$, for the $\{m,n\}$
summations. The normalization coefficient, $(\Delta r)^2/(2\pi)$,  has been
introduced to make $F$ similar to a Fourier integral at the limit,
$\Delta r \rightarrow 0$.

A similar comment to the one-dimensional case is in order for the radial
case. An approximate reconstruction of the Fourier transform function
appears to be allowed following formula \eqref{approxfourier1}

Figure \ref{7} shows that, for a finite value of $\Delta r$, hence for finite
summations governed by the resulting $K$, negative values of $F$ can be
found,
even for a ``rebel case'' like that shown in the right part of Fig. \ref{6}. 
A search
for negative values of $F$ under moderate values of $\Delta r$ returns 
a detection rate of $\sim 90\%$ through our set of 9884 functions. This is
significantly better than the result discussed in the previous subsection, 
where the practical tests based on ``Bochner method'' appears to be only 
very slowly evolving 
with already high matrix order.

 Fig.\ref{7} calls for a comment. The plots appear to approximately satisfy
 a radial symmetry in the two-dimensional $(\al,\gamma)$ plane, while the
 resummation formula \eqref{eqradial} is not {\it a priori} symmetric. The reason
 is that it gives an approximate reconstruction of the radial function $\va(s)$
 as predicted by our general argument of section II for the radial 2-dimensional
 case. This reconstruction property with a finite number of terms in 
\eqref{eqradial} is exemplified in Fig.\ref{8} with the non Fourier-positive 
function depicted in Fig.\ref{5}. The negative domain is detected already with 
$K=40$ and reasonably fully reproduced with $K=80.$

\begin{figure}
\scalebox{.65}{\includegraphics{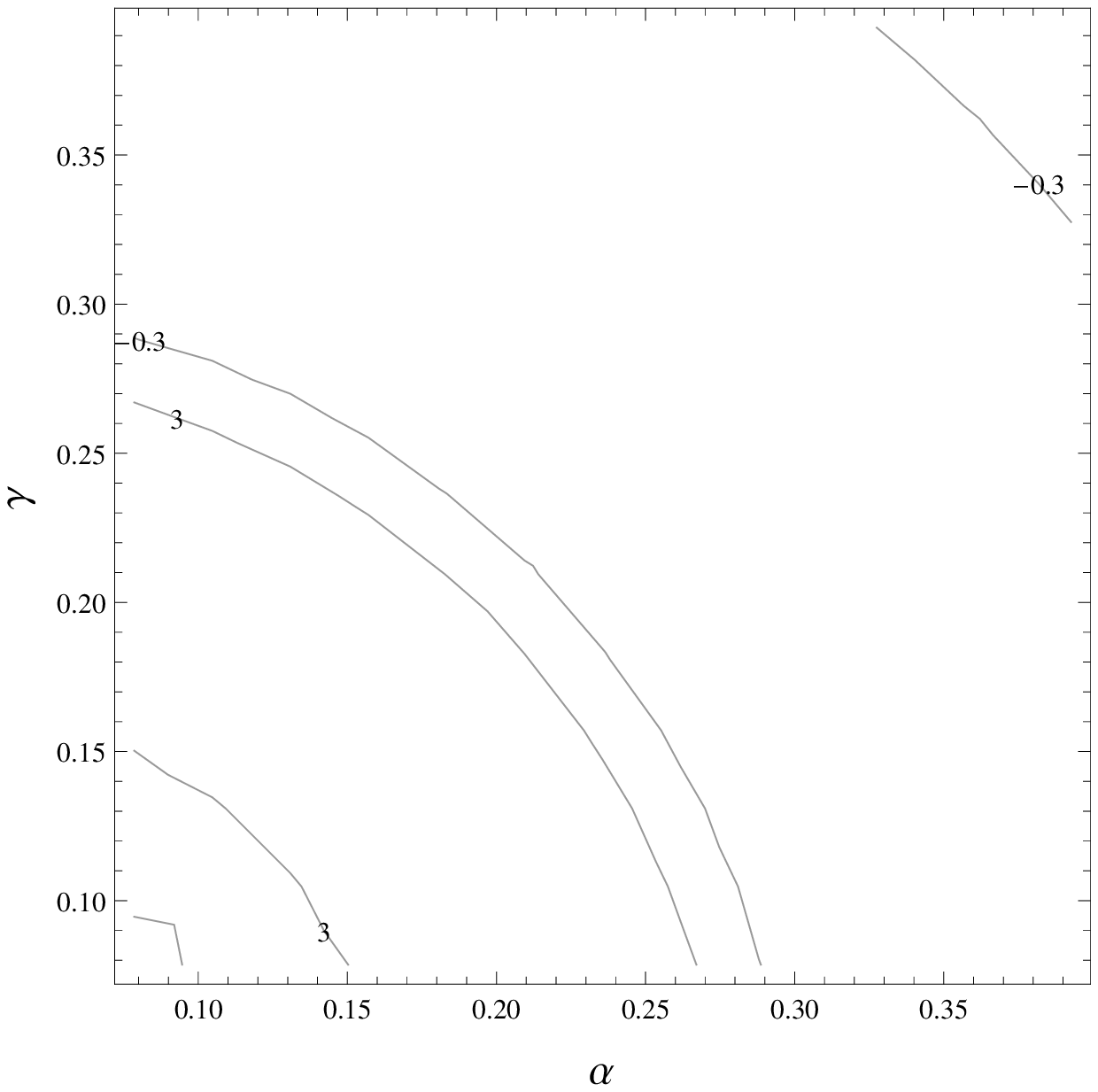}}
\scalebox{.65}{\includegraphics{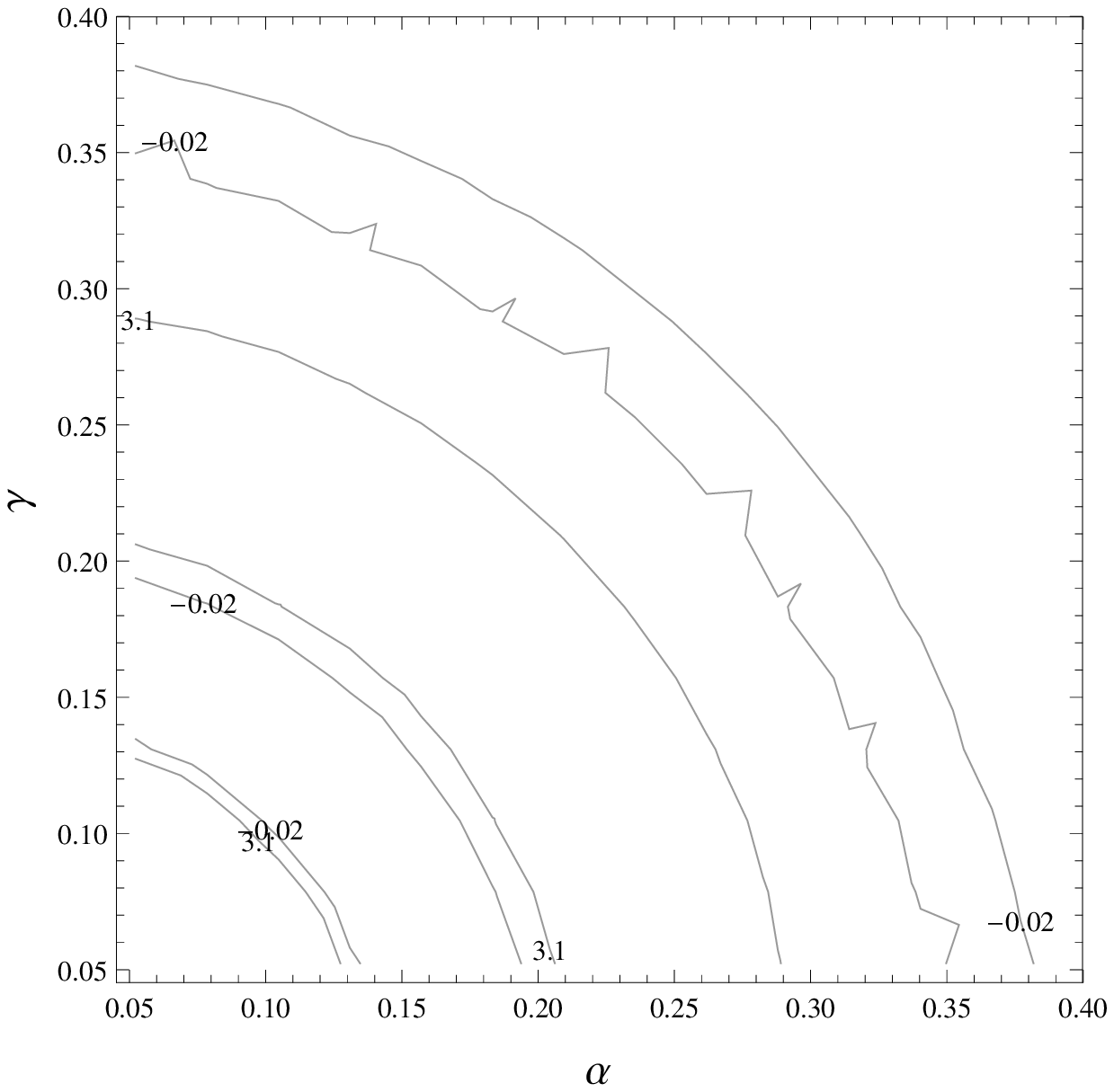}}
\caption{{\it Contour plot of $F(\alpha,\gamma)$ when $\Delta r=.5$.} Left:
same
case as the ``success'' case shown in the left part of Fig.\ref{6};
contours for
$F=-.3,3$ are shown. Right: same case as the ``failure'' case shown in the
right part of Fig.\ref{6}; but now, a contour with a negative value,
$F=-.02$,
provides a detection.}
\la{7}
\end{figure}

\begin{figure}
\scalebox{.65}{\includegraphics{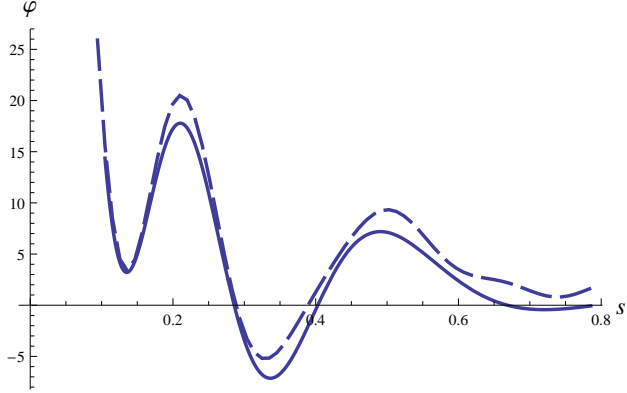}}
\caption{{\it Example of  a ``reconstruction'' of a radial
non-Fourier-positive test function.} The function $\varphi_{pn}$ 
depicted in the right part of Fig.\ref{5} by a dashed line
is approximated respectively with 
$K=40$ (large dashes, detecting one negative domain) and with $K=80$
(full line, detecting both negative domains).}  
\la{8}
\end{figure}

\section{Summary and outlook}
In any dimension $d$ there exists a set of real functions,
partners under Fourier transform, that are both positive. We have called them {\it Fourier-positive functions}. If only for purely
mathematical curiosity, the mapping between these two convex, partner sets
is of some interest. But this interest is reinforced by the fact that, in
theoretical physics, it happens that a ``density'' in one space has a Fourier
partner which is also a ``density''. (By definition, densities are positive
 observables.)

There is also an interesting ``subproblem'', that of the mapping between
subsets, for instance polynomials multiplied by Gaussians in both spaces,
or polynomials multiplied by simple exponentials in one space with rational
function partners. Such subsets are nested according to the order of
the considered polynomials. This hierarchy allows a useful set of successive
approximations.

In fact a general, both necessary and sufficient, mathematical 
criterion for ensuring
 Fourier-positivity seems  not to be yet known. Indeed, examples taken 
from physics show that
Fourier-positivity is a nontrivial  constraint on density models \cite{lappi},
where small modifications may play a role.
For a simple illustration, compare a  square well density, whose 
Fourier transform is 
the Bessel function, with a Gaussian density, which, being 
invariant by Fourier transform, is obviously Fourier positive.

Modern computers allow ``Fast Fourier Transforms'' which give an easy answer
to positivity properties in both partner spaces, but, obviously, a pure
numerical approach is not completely satisfactory. The present work gives
several mathematical, analytical arguments to complement our previous work 
\cite{gipe}, where it was shown that the topology of such interesting 
``positive partner subsets'' was highly non trivial and moreover, where
there appeared an intuition that extremal elements in such convex sets are
reminiscent of Dirac combs.

The proofs displayed in this work do take advantage of this intuition, but
indirectly. In substance, we use two kinds of criteria to test whether a
positive
``object'' $\psi$ has a positive ``image'' $\varphi$, criteria that use 
values of $\psi$ only. i) A positivy criterion for the characteristic Poisson
function associated with $\psi$ through a ``Dirac comb'' distribution 
turns out to be quite efficient for both $d=1$
and radial, $d=2$ cases. ii) A criterion taken from Boechner's theorem,
namely the positivity of Toeplitz matrices and similar matrices, turns out
to be efficient for $d=1$ cases, but disappointing for radial, d=2 cases.
In both cases, we took great care to validate our analytical considerations
by means of controled, numerical tests, including statistical evaluations.

A third set of criteria \cite{newfourier} was in fact our initial approach.
It consisted in relating positivity of a function and convexity of analytical
continuations of related functions, to define bounds via the Jensen's theorem,
but we failed in making this approach a convincing one, if only because
analytical continuation most often has to face severe singularities. It is
not displayed in the present work. We keep it on a back-burner.

On a deeper but difficult level, the question of a general criterion of Fourier
 positivity and a classification of those functions still remains widely open. 
We hope that the consideration of Dirac combs and Poisson resummation proposed 
in our paper in this context may help to make some new steps in that problem.

For a more immediate outlook, our priority will be to take advantage 
of the hierarchy of
mapped subspaces abovementioned and use it for actual physical problems. 
Reliable error bars are essential in data analysis and we want to use our
criteria for both estimates of $\psi$ and $\varphi$.

\section{Acknowledgements}
We want to thank Bertrand Eynard, Philippe Jaming, Jean-Pierre Kahane  and Cyrille Marquet for stimulating discussions.

\end{document}